\input harvmac
\input epsf

%
\let\includefigures=\iftrue
%
%
%
\newfam\black
\input rotate
\input epsf
\noblackbox
%
%
\includefigures
\message{If you do not have epsf.tex (to include figures),}
\message{change the option at the top of the tex file.}
\def\figin{\epsfcheck\figin}\def\figins{\epsfcheck\figins}
\def\epsfcheck{\ifx\epsfbox\UnDeFiNeD
\message{(NO epsf.tex, FIGURES WILL BE IGNORED)}
\gdef\figin##1{\vskip2in}\gdef\figins##1{\hskip.5in}
\else\message{(FIGURES WILL BE INCLUDED)}%
\gdef\figin##1{##1}\gdef\figins##1{##1}\fi}
\def\DefWarn#1{}
\def\N{{\cal N}}
\def\figinsert{\goodbreak\midinsert}
\def\ifig#1#2#3{\DefWarn#1\xdef#1{fig.~\the\figno}
\writedef{#1\leftbracket fig.\noexpand~\the\figno}%
\figinsert\figin{\centerline{#3}}\medskip\centerline{\vbox{\baselineskip12pt
\advance\hsize by -1truein\noindent\footnotefont{\bf
Fig.~\the\figno:} #2}}
\bigskip\endinsert\global\advance\figno by1}
\else
\def\ifig#1#2#3{\xdef#1{fig.~\the\figno}
\writedef{#1\leftbracket fig.\noexpand~\the\figno}%
\global\advance\figno by1} \fi

\def\tilde{\widetilde}

\def\subsubsec#1{\bigskip\noindent{\it #1}}
\def\yboxit#1#2{\vbox{\hrule height #1 \hbox{\vrule width #1
\vbox{#2}\vrule width #1 }\hrule height #1 }}
\def\fillbox#1{\hbox to #1{\vbox to #1{\vfil}\hfil}}
\def\ybox{{\lower 1.3pt \yboxit{0.4pt}{\fillbox{8pt}}\hskip-0.2pt}}

\def\rightarrowbox#1#2{
  \setbox1=\hbox{\kern#1{${ #2}$}\kern#1}
  \,\vbox{\offinterlineskip\hbox to\wd1{\hfil\copy1\hfil}
    \kern 3pt\hbox to\wd1{\rightarrowfill}}}

\def\QR{\Bbb{R}}

\def\QZ{\Bbb{Z}}

\def\half{{1\over 2}}
\def\Tr{{{\rm Tr~ }}}

\def\Im{{\rm Im\hskip0.1em}}

\def\ket#1{|#1\rangle}

\def\vev#1{\langle{#1}\rangle}

\def\CA{{\cal A}}

\def\CO{{\cal O}}
\def\O{{\cal O}}

\def\tilde{\widetilde}

\def\II{\relax{I\kern-.10em I}}

\def\bar{\overline}

\def\IZ{\relax\ifmmode\mathchoice
{\hbox{\cmss Z\kern-.4em Z}}{\hbox{\cmss Z\kern-.4em Z}}
{\lower.9pt\hbox{\cmsss Z\kern-.4em Z}} {\lower1.2pt\hbox{\cmsss
Z\kern-.4em Z}}\else{\cmss Z\kern-.4em Z}\fi}
\def\IB{\relax{\rm I\kern-.18em B}}
\def\IC{{\relax\hbox{$\inbar\kern-.3em{\rm C}$}}}
\def\ID{\relax{\rm I\kern-.18em D}}
\def\IE{\relax{\rm I\kern-.18em E}}
\def\IF{\relax{\rm I\kern-.18em F}}
\def\IG{\relax\hbox{$\inbar\kern-.3em{\rm G}$}}
\def\IGa{\relax\hbox{${\rm I}\kern-.18em\Gamma$}}
\def\IH{\relax{\rm I\kern-.18em H}}
\def\II{\relax{\rm I\kern-.18em I}}
\def\IK{\relax{\rm I\kern-.18em K}}
\def\IN{\relax{\rm I\kern-.18em N}}
\def\IP{\relax{\rm I\kern-.18em P}}

%
\def\inbar{\,\vrule height1.5ex width.4pt depth0pt}

\font\cmss=cmss10 \font\cmsss=cmss10 at 7pt
\def\IR{\relax{\rm I\kern-.18em R}}

\def\lp10{l_P^{10}}
\def\lp11{l_P^{11}}
\def\R11{R_{11}}

\def\lt{\tilde\lambda}

\def\gb#1{ {\langle #1 ] } }

\def\K{t}

\def\D{\dot}
\def\la{\lambda}
\def\a{\alpha}
\def\b{\beta}
\def\braket{\vev}
\def\W{\widetilde}

\newbox\tmpbox\setbox\tmpbox\hbox{\abstractfont
}
 \Title{\vbox{\baselineskip12pt\hbox to\wd\tmpbox{\hss
 hep-ph/0503132} }}
 {\vbox{\centerline{One-Loop Amplitudes Of Gluons In SQCD}
 \bigskip
 \centerline{}
 }}
\smallskip
\centerline{Ruth Britto, Evgeny Buchbinder, Freddy Cachazo, Bo
Feng}
\smallskip
\bigskip
\centerline{\it School of Natural Sciences, Institute for Advanced
Study, Princeton NJ 08540 USA}
\bigskip
\vskip 1cm \noindent

\input amssym.tex

One-loop amplitudes of gluons in supersymmetric Yang-Mills are
four-dimensional cut-constructible. This means that they can be
determined from their unitarity cuts. We present a new systematic
procedure to explicitly carry out any finite unitarity cut
integral. The procedure naturally separates the contributions from
bubble, triangle and box scalar integrals. This technique allows
the systematic calculation of ${\cal N}=1$ amplitudes of gluons.
As an application we compute all next-to-MHV six-gluon amplitudes
in ${\cal N}=1$ super-Yang-Mills.

\Date{March 2005}
%

\lref\cuts{ L.D. Landau, Nucl.\ Phys. {\bf 13}, 181 (1959);
 S. Mandelstam, Phys. Rev. {\bf 112}, 1344 (1958), {\bf 115}, 1741 (1959);
 R.E. Cutskosky, J. Math. Phys. {\bf 1}, 429 (1960).
}

\lref\bdknmhv{
 Z.~Bern, L.~J.~Dixon and D.~A.~Kosower,
``All Next-to-Maximally-Helicity-Violating One-Loop Gluon Amplitudes in N=4 Super-Yang-Mills Theory,'' hep-th/0412210.}

\lref\DennerQQ{
A.~Denner, U.~Nierste and R.~Scharf,
``A Compact expression for the scalar one loop four point function,''
Nucl.\ Phys.\ B {\bf 367}, 637 (1991).
}

\lref\BernZX{
  Z.~Bern, L.~J.~Dixon, D.~C.~Dunbar and D.~A.~Kosower,
  ``One loop n point gauge theory amplitudes, unitarity and collinear limits,''
  Nucl.\ Phys.\ B {\bf 425}, 217 (1994)
  [arXiv:hep-ph/9403226].
}

\lref\BernCG{
  Z.~Bern, L.~J.~Dixon, D.~C.~Dunbar and D.~A.~Kosower,
  ``Fusing gauge theory tree amplitudes into loop amplitudes,''
  Nucl.\ Phys.\ B {\bf 435}, 59 (1995)
  [arXiv:hep-ph/9409265].
}

\lref\WittenNN{
E.~Witten,
``Perturbative gauge theory as a string theory in twistor space,''
Commun.\ Math.\ Phys.\  {\bf 252}, 189 (2004)
[arXiv:hep-th/0312171].
}

\lref\CachazoKJ{
F.~Cachazo, P.~Svr\v{c}ek and E.~Witten,
``MHV vertices and tree amplitudes in gauge theory,''
JHEP {\bf 0409}, 006 (2004)
[arXiv:hep-th/0403047].
}

\lref\BerkovitsJJ{
N.~Berkovits and E.~Witten,
``Conformal supergravity in twistor-string theory,''
JHEP {\bf 0408}, 009 (2004)
[arXiv:hep-th/0406051].
}

\lref\penrose{R. Penrose, ``Twistor Algebra,'' J. Math. Phys. {\bf
8} (1967) 345.}

\lref\berends{F. A. Berends, W. T. Giele and H. Kuijf, ``On
Relations Between Multi-Gluon And Multi-Graviton Scattering,"
Phys. Lett {\bf B211} (1988) 91.}

\lref\berendsgluon{F. A. Berends, W. T. Giele and H. Kuijf,
``Exact and Approximate Expressions for Multigluon Scattering,"
Nucl. Phys. {\bf B333} (1990) 120.}

\lref\bernplusa{Z. Bern, L. Dixon and D. A. Kosower, ``New QCD
Results From String Theory,'' in {\it Strings '93}, ed. M. B.
Halpern et. al. (World-Scientific, 1995), hep-th/9311026.}

\lref\bernplusb{Z. Bern, G. Chalmers, L. J. Dixon and D. A.
Kosower, ``One Loop $N$ Gluon Amplitudes with Maximal Helicity
Violation via Collinear Limits," Phys. Rev. Lett. {\bf 72} (1994)
2134.}

\lref\bernfive{Z. Bern, L. J. Dixon and D. A. Kosower, ``One Loop
Corrections to Five Gluon Amplitudes," Phys. Rev. Lett {\bf 70}
(1993) 2677.}

\lref\bernfourqcd{Z.Bern and  D. A. Kosower, "The Computation of
Loop Amplitudes in Gauge Theories," Nucl. Phys.  {\bf B379,}
(1992) 451.}

\lref\cremmerlag{E. Cremmer and B. Julia, ``The $N=8$ Supergravity
Theory. I. The Lagrangian," Phys. Lett.  {\bf B80} (1980) 48.}

\lref\cremmerso{E. Cremmer and B. Julia, ``The $SO(8)$
Supergravity," Nucl. Phys.  {\bf B159} (1979) 141.}

\lref\dewitt{B. DeWitt, "Quantum Theory of Gravity, III:
Applications of Covariant Theory," Phys. Rev. {\bf 162} (1967)
1239.}

\lref\dunbarn{D. C. Dunbar and P. S. Norridge, "Calculation of
Graviton Scattering Amplitudes Using String Based Methods," Nucl.
Phys. B {\bf 433,} 181 (1995), hep-th/9408014.}

\lref\ellissexton{R. K. Ellis and J. C. Sexton, "QCD Radiative
corrections to parton-parton scattering," Nucl. Phys.  {\bf B269}
(1986) 445.}

\lref\gravityloops{Z. Bern, L. Dixon, M. Perelstein, and J. S.
Rozowsky, ``Multi-Leg One-Loop Gravity Amplitudes from Gauge
Theory,"  hep-th/9811140.}

\lref\kunsztqcd{Z. Kunszt, A. Signer and Z. Tr\'{o}cs\'{a}nyi,
``One-loop Helicity Amplitudes For All $2\rightarrow2$ Processes
in QCD and ${\cal N}=1$ Supersymmetric Yang-Mills Theory,'' Nucl.
Phys.  {\bf B411} (1994) 397, hep-th/9305239.}

\lref\mahlona{G. Mahlon, ``One Loop Multi-photon Helicity
Amplitudes,'' Phys. Rev.  {\bf D49} (1994) 2197, hep-th/9311213.}

\lref\mahlonb{G. Mahlon, ``Multi-gluon Helicity Amplitudes
Involving a Quark Loop,''  Phys. Rev.  {\bf D49} (1994) 4438,
hep-th/9312276.}

\lref\klt{H. Kawai, D. C. Lewellen and S.-H. H. Tye, ``A Relation
Between Tree Amplitudes of Closed and Open Strings," Nucl. Phys.
{B269} (1986) 1.}

\lref\pppmgr{Z. Bern, D. C. Dunbar and T. Shimada, ``String Based
Methods In Perturbative Gravity," Phys. Lett.  {\bf B312} (1993)
277, hep-th/9307001.}

\lref\GiombiIX{ S.~Giombi, R.~Ricci, D.~Robles-Llana and
D.~Trancanelli, ``A Note on Twistor Gravity Amplitudes,''
hep-th/0405086.
}

\lref\sbook{
R. J. Eden, P. V. Landshoff, D. I. Olive and J. C. Polkinghorne,
{\it The Analytic S-Matrix}, Cambridge University Press, 1966.
}

\lref\WuFB{
  J.~B.~Wu and C.~J.~Zhu,
  ``MHV vertices and scattering amplitudes in gauge theory,''
  JHEP {\bf 0407}, 032 (2004)
  [arXiv:hep-th/0406085].
}

\lref\Feynman{R.P. Feynman, Acta Phys. Pol. 24 (1963) 697, and in
{\it Magic Without Magic}, ed. J. R. Klauder (Freeman, New York,
1972), p. 355.}

\lref\Peskin{M.~E. Peskin and D.~V. Schroeder, {\it An Introduction
to Quantum Field Theory} (Addison-Wesley Pub. Co., 1995).}

\lref\parke{S. Parke and T. Taylor, ``An Amplitude For $N$ Gluon
Scattering,'' Phys. Rev. Lett. {\bf 56} (1986) 2459; F. A. Berends
and W. T. Giele, ``Recursive Calculations For Processes With $N$
Gluons,'' Nucl. Phys. {\bf B306} (1988) 759. }

\lref\BrandhuberYW{ A.~Brandhuber, B.~Spence and G.~Travaglini,
``One-Loop Gauge Theory Amplitudes In N = 4 Super Yang-Mills From
MHV Vertices,'' hep-th/0407214.
}

\lref\loopmhv{A.~Brandhuber, B.~Spence and G.~Travaglini,
``One-Loop Gauge Theory Amplitudes In N = 4 Super Yang-Mills From
MHV Vertices,'' Nucl.\ Phys.\ B {\bf 706}, 150 (2005)
  [arXiv:hep-th/0407214];
C.~Quigley and M.~Rozali,
  ``One-loop MHV amplitudes in supersymmetric gauge theories,''
  JHEP {\bf 0501}, 053 (2005)
  [arXiv:hep-th/0410278];
J.~Bedford, A.~Brandhuber, B.~Spence and G.~Travaglini,
  ``A twistor approach to one-loop amplitudes in N = 1 supersymmetric
  Yang-Mills theory,''
  Nucl.\ Phys.\ B {\bf 706}, 100 (2005)
  [arXiv:hep-th/0410280].
}

\lref\CachazoZB{
F.~Cachazo, P.~Svr\v{c}ek and E.~Witten,
``Twistor space structure of one-loop amplitudes in gauge theory,''
JHEP {\bf 0410}, 074 (2004)
[arXiv:hep-th/0406177].
}

\lref\passarino{ L.~M. Brown and R.~P. Feynman, ``Radiative Corrections To Compton Scattering,'' Phys. Rev. 85:231
(1952); G.~Passarino and M.~Veltman, ``One Loop Corrections For E+ E- Annihilation Into Mu+ Mu- In The Weinberg
Model,'' Nucl. Phys. B160:151 (1979);
G.~'t Hooft and M.~Veltman, ``Scalar One Loop Integrals,'' Nucl. Phys. B153:365 (1979); R.~G.~
Stuart, ``Algebraic Reduction Of One Loop Feynman Diagrams To Scalar Integrals,'' Comp. Phys. Comm. 48:367 (1988); R.~G.~Stuart and A.~Gongora, ``Algebraic Reduction Of One Loop Feynman Diagrams To Scalar Integrals. 2,'' Comp. Phys. Comm. 56:337 (1990).}

\lref\neerven{ W. van Neerven and J. A. M. Vermaseren, ``Large Loop Integrals,'' Phys. Lett.
137B:241 (1984)}

\lref\melrose{ D.~B.~Melrose, ``Reduction Of Feynman Diagrams,'' Il Nuovo Cimento 40A:181 (1965); G.~J.~van Oldenborgh and J.~A.~M.~Vermaseren, ``New Algorithms For One Loop Integrals,'' Z. Phys. C46:425 (1990);
G.J. van Oldenborgh,  PhD Thesis, University of Amsterdam (1990);
A. Aeppli, PhD thesis, University of Zurich (1992).}

\lref\bernTasi{Z.~Bern, hep-ph/9304249, in {\it Proceedings of
Theoretical Advanced Study Institute in High Energy Physics (TASI
92)}, eds. J. Harvey and J. Polchinski (World Scientific, 1993). }

\lref\morgan{ Z.~Bern and A.~G.~Morgan, ``Supersymmetry relations
between contributions to one loop gauge boson amplitudes,'' Phys.\
Rev.\ D {\bf 49}, 6155 (1994), hep-ph/9312218.
}

\lref\RoiSpV{R.~Roiban, M.~Spradlin and A.~Volovich, ``A Googly
Amplitude From The B-Model In Twistor Space,'' JHEP {\bf 0404},
012 (2004) hep-th/0402016; R.~Roiban and A.~Volovich, ``All Googly
Amplitudes From The $B$-Model In Twistor Space,''Phys.\ Rev.\ Lett.\  {\bf 93}, 131602 (2004)
[arXiv:hep-th/0402121];
R.~Roiban, M.~Spradlin and A.~Volovich, ``On The Tree-Level
S-Matrix Of Yang-Mills Theory,'' Phys.\ Rev.\ D {\bf 70}, 026009
(2004) hep-th/0403190,
S.~Gukov, L.~Motl and A.~Neitzke,
``Equivalence of twistor prescriptions for super Yang-Mills,''
arXiv:hep-th/0404085,
I.~Bena, Z.~Bern and D.~A.~Kosower,
``Twistor-space recursive formulation of gauge theory amplitudes,''
arXiv:hep-th/0406133.
}

\lref\CachazoBY{
  F.~Cachazo, P.~Svrcek and E.~Witten,
  ``Gauge theory amplitudes in twistor space and holomorphic anomaly,''
  JHEP {\bf 0410}, 077 (2004)
  [arXiv:hep-th/0409245].
}

\lref\DixonWI{ L.~J.~Dixon, ``Calculating Scattering Amplitudes
Efficiently,'' hep-ph/9601359.
}

\lref\BernMQ{ Z.~Bern, L.~J.~Dixon and D.~A.~Kosower, ``One Loop
Corrections To Five Gluon Amplitudes,'' Phys.\ Rev.\ Lett.\  {\bf
70}, 2677 (1993), hep-ph/9302280.
}

\lref\berends{F.~A.~Berends, R.~Kleiss, P.~De Causmaecker, R.~Gastmans and T.~T.~Wu, ``Single Bremsstrahlung Processes In Gauge Theories,'' Phys. Lett. {\bf B103} (1981) 124; P.~De
Causmaeker, R.~Gastmans, W.~Troost and T.~T.~Wu, ``Multiple Bremsstrahlung In Gauge Theories At High-Energies. 1. General
Formalism For Quantum Electrodynamics,'' Nucl. Phys. {\bf
B206} (1982) 53; R.~Kleiss and W.~J.~Stirling, ``Spinor Techniques For Calculating P Anti-P $\to$ W+- / Z0 + Jets,'' Nucl. Phys. {\bf
B262} (1985) 235; R.~Gastmans and T.~T. Wu, {\it The Ubiquitous
Photon: Heliclity Method For QED And QCD} Clarendon Press, 1990.}

\lref\xu{Z. Xu, D.-H. Zhang and L. Chang, ``Helicity Amplitudes For Multiple
Bremsstrahlung In Massless Nonabelian Theories,''
 Nucl. Phys. {\bf B291}
(1987) 392.}

\lref\gunion{J.~F. Gunion and Z. Kunszt, ``Improved Analytic Techniques For Tree Graph Calculations And The G G Q
Anti-Q Lepton Anti-Lepton Subprocess,''
Phys. Lett. {\bf 161B}
(1985) 333.}

\lref\GeorgiouBY{ G.~Georgiou, E.~W.~N.~Glover and V.~V.~Khoze,
``Non-MHV Tree Amplitudes In Gauge Theory,'' JHEP {\bf 0407}, 048
(2004) [arXiv:hep-th/0407027].
}

\lref\WuJX{
  J.~B.~Wu and C.~J.~Zhu,
  ``MHV vertices and fermionic scattering amplitudes in gauge theory with
  quarks and gluinos,''
  JHEP {\bf 0409}, 063 (2004)
  [arXiv:hep-th/0406146].
}

\lref\WuFB{ J.~B.~Wu and C.~J.~Zhu, ``MHV Vertices And Scattering
Amplitudes In Gauge Theory,'' JHEP {\bf 0407}, 032 (2004)
[arXiv:hep-th/0406085].
}

\lref\GeorgiouWU{ G.~Georgiou and V.~V.~Khoze, ``Tree Amplitudes
In Gauge Theory As Scalar MHV Diagrams,'' JHEP {\bf 0405}, 070
(2004) [arXiv:hep-th/0404072].
}

\lref\Nair{V. Nair, ``A Current Algebra For Some Gauge Theory
Amplitudes," Phys. Lett. {\bf B78} (1978) 464. }

\lref\BernAD{ Z.~Bern, ``String Based Perturbative Methods For
Gauge Theories,'' hep-ph/9304249.
}

\lref\BernKR{ Z.~Bern, L.~J.~Dixon and D.~A.~Kosower,
``Dimensionally Regulated Pentagon Integrals,'' Nucl.\ Phys.\ B
{\bf 412}, 751 (1994), hep-ph/9306240.
}

\lref\CachazoDR{ F.~Cachazo, ``Holomorphic Anomaly Of Unitarity
Cuts And One-Loop Gauge Theory Amplitudes,'' hep-th/0410077.
}

\lref\giel{W. T. Giele and E. W. N. Glover, ``Higher order corrections to jet cross-sections in e+ e- annihilation,'' Phys. Rev. {\bf D46}
(1992) 1980; W. T. Giele, E. W. N. Glover and D. A. Kosower, ``Higher order corrections to jet cross-sections in hadron colliders,'' Nucl.
Phys. {\bf B403} (1993) 633. }

\lref\kuni{Z. Kunszt and D. Soper, ``Calculation of jet cross-sections in hadron collisions at order alpha-s**3,''Phys. Rev. {\bf D46} (1992)
192; Z. Kunszt, A. Signer and Z. Tr\' ocs\' anyi, ``Singular terms of helicity amplitudes at one loop in QCD and the soft limit
of the cross-sections of multiparton processes,'' Nucl. Phys. {\bf
B420} (1994) 550. }

\lref\seventree{F.~A. Berends, W.~T. Giele and H. Kuijf, ``Exact And Approximate Expressions For Multi - Gluon Scattering,'' Nucl. Phys.
{\bf B333} (1990) 120.}

\lref\mangpxu{M. Mangano, S.~J. Parke and Z. Xu, ``Duality And Multi - Gluon Scattering,'' Nucl. Phys. {\bf B298}
(1988) 653.}

\lref\colorord{ Z.~Bern and D.~A.~Kosower,
  ``Color Decomposition Of One Loop Amplitudes In Gauge Theories,''
  Nucl.\ Phys.\ B {\bf 362}, 389 (1991); F.~A.~Berends and W.~Giele,
  ``The Six Gluon Process As An Example Of Weyl-Van Der Waerden Spinor
  Calculus,''
  Nucl.\ Phys.\ B {\bf 294}, 700 (1987); M. Mangano, S.~J. Parke and Z. Xu, ``Duality And Multi - Gluon Scattering,'' Nucl. Phys. {\bf B298}
(1988) 653; M.~L.~Mangano,
  ``The Color Structure Of Gluon Emission,''
  Nucl.\ Phys.\ B {\bf 309}, 461 (1988);}

\lref\mangparke{M. Mangano and S.~J. Parke, ``Multiparton Amplitudes In Gauge Theories,'' Phys. Rep. {\bf 200}
(1991) 301.}

\lref\grisaru{M. T. Grisaru, H. N. Pendleton and P. van Nieuwenhuizen, ``Supergravity And The S Matrix,'' Phys. Rev.  {\bf D15} (1977) 996; M. T. Grisaru and H. N. Pendleton, ``Some Properties Of Scattering Amplitudes In Supersymmetric Theories,'' Nucl. Phys. {\bf B124} (1977) 81.}

\lref\Bena{I. Bena, Z. Bern, D. A. Kosower and R. Roiban, ``Loops in Twistor Space,'' hep-th/0410054.}

\lref\BernKY{
Z.~Bern, V.~Del Duca, L.~J.~Dixon and D.~A.~Kosower,
``All Non-Maximally-Helicity-Violating One-Loop Seven-Gluon Amplitudes In N =
4 Super-Yang-Mills Theory,''
arXiv:hep-th/0410224.
}

\lref\BrittoNJ{
  R.~Britto, F.~Cachazo and B.~Feng,
  ``Computing one-loop amplitudes from the holomorphic anomaly of unitarity
  cuts,''
  Phys.\ Rev.\ D {\bf 71}, 025012 (2005)
  [arXiv:hep-th/0410179].
}

\lref\BidderVX{
  S.~J.~Bidder, N.~E.~J.~Bjerrum-Bohr, D.~C.~Dunbar and W.~B.~Perkins,
  ``Twistor space structure of the box coefficients of N = 1 one-loop
  amplitudes,''
  arXiv:hep-th/0412023.
}

\lref\BidderTX{
S.~J.~Bidder, N.~E.~J.~Bjerrum-Bohr, L.~J.~Dixon and D.~C.~Dunbar,
``N = 1 supersymmetric one-loop amplitudes and the holomorphic anomaly of
unitarity cuts,''
arXiv:hep-th/0410296.
}

\lref\BrittoNC{
R.~Britto, F.~Cachazo and B.~Feng,
``Generalized unitarity and one-loop amplitudes in N = 4 super-Yang-Mills,''
arXiv:hep-th/0412103.
}

\lref\BernIX{
Z.~Bern and G.~Chalmers,
``Factorization in one loop gauge theory,''
Nucl.\ Phys.\ B {\bf 447}, 465 (1995)
[arXiv:hep-ph/9503236].
}

\lref\LuoMY{
M.~x.~Luo and C.~k.~Wen,
``Compact formulas for all tree amplitudes of six partons,''
arXiv:hep-th/0502009.
}

\lref\LuoRX{
M.~x.~Luo and C.~k.~Wen,
``Recursion relations for tree amplitudes in super gauge theories,''
arXiv:hep-th/0501121.
}

\lref\BrittoAP{
R.~Britto, F.~Cachazo and B.~Feng,
``New recursion relations for tree amplitudes of gluons,''
arXiv:hep-th/0412308.
}
\lref\BidderRI{
  S.~J.~Bidder, N.~E.~J.~Bjerrum-Bohr, D.~C.~Dunbar and W.~B.~Perkins,
  ``One-loop gluon scattering amplitudes in theories with N $<$ 4
  supersymmetries,''
  arXiv:hep-th/0502028.
}

\lref\BernJE{
  Z.~Bern, L.~J.~Dixon and D.~A.~Kosower,
  ``Progress in one-loop QCD computations,''
  Ann.\ Rev.\ Nucl.\ Part.\ Sci.\  {\bf 46}, 109 (1996)
  [arXiv:hep-ph/9602280].
}

\lref\BedfordNH{
  J.~Bedford, A.~Brandhuber, B.~Spence and G.~Travaglini,
  ``Non-supersymmetric loop amplitudes and MHV vertices,''
  hep-th/0412108.
}

\lref\MahlonSI{
  G.~Mahlon,
  ``Multi - gluon helicity amplitudes involving a quark loop,''
  Phys.\ Rev.\ D {\bf 49}, 4438 (1994)
  [arXiv:hep-ph/9312276].
}

\lref\BrittoFQ{
  R.~Britto, F.~Cachazo, B.~Feng and E.~Witten,
  ``Direct proof of tree-level recursion relation in Yang-Mills theory,''
  hep-th/0501052.
}


\lref\BernSC{
  Z.~Bern, L.~J.~Dixon and D.~A.~Kosower,
  ``One-loop amplitudes for e+ e- to four partons,''
  Nucl.\ Phys.\ B {\bf 513}, 3 (1998)
  [arXiv:hep-ph/9708239].
}

\lref\BernDN{
  Z.~Bern, L.~J.~Dixon and D.~A.~Kosower,
  ``A two-loop four-gluon helicity amplitude in QCD,''
  JHEP {\bf 0001}, 027 (2000)
  [arXiv:hep-ph/0001001].
}

\lref\BernDB{
  Z.~Bern and A.~G.~Morgan,
  ``Massive Loop Amplitudes from Unitarity,''
  Nucl.\ Phys.\ B {\bf 467}, 479 (1996)
  [arXiv:hep-ph/9511336].
}


\newsec{Introduction}

Scattering amplitudes of gluons in Yang-Mills theories exhibit a
remarkable simplicity that is not manifest from their calculation
using Feynman diagrams. At tree-level, Parke-Taylor or maximally
helicity violating (MHV) amplitudes
\parke\ provide a striking example.

One-loop MHV amplitudes in  ${\cal N}=4$ super Yang-Mills also
exhibit remarkable simplicity when computed using the unitarity
based method \refs{\BernZX, \BernCG,\BernDB}. Recently, a new
technique was presented for computing general one-loop amplitudes in
${\cal N}=4$ super Yang-Mills using quadruple cuts \BrittoNC. This
is a systematic and simple procedure that blends old
\refs{\Feynman,\sbook}, more modern \refs{\BernZX, \BernCG, \BernDB, \BernSC,\BernDN,\BernKY}
and very recent ideas \refs{\WittenNN,\CachazoZB} to uncover the
simplicity of generic amplitudes. Using this one can easily
reproduce all known results \refs{\BernZX, \BernCG, \CachazoDR,
\BrittoNJ, \BernKY, \bdknmhv} and in principle compute any other
amplitude.

One main motivation for computing ${\cal N}=4$ amplitudes of
gluons is that they are part of amplitudes in theories with less
supersymmetry \BernZX\ (for a review, see \DixonWI).

In this paper we concentrate on one-loop $\N =1$ amplitudes of gluons. Such
an amplitude can be decomposed as follows,
\eqn\sipunc{\CA^{\N=1~{\rm vector}}=\CA^{\N=4}-3\CA^{\N=1} }
where $\CA^{\N=4}$ is an amplitude where the full $\N=4$ multiplet
runs in the loop, and  $\CA^{\N=1}$ denotes the contribution from
an $\N=1$ chiral supermultiplet running in the loop. As mentioned
above, the $\N=4$ problem is easy to solve using quadruple cuts.
On the other hand, $\CA^{\N=1}$ only contains fermions and scalars
in the loop and thus it is expected to be simpler than the full
$\N=1$ vector multiplet. This is why the decomposition \sipunc\ is
useful.

The computation of $\CA^{\N=1}$ is also important because it is
part of a supersymmetry decomposition of a QCD amplitude at next-to-leading order,
\eqn\themiste{\CA^{\rm QCD}=\CA^{\N=4}-4\CA^{\N=1}+\CA^{\rm
scalar} }
where $\CA^{\rm QCD}$ denotes an amplitude with only a gluon
running in the loop. $\CA^{\rm scalar}$ is an amplitude with only
a complex scalar running in the loop.

The benefit of this approach is that supersymmetric amplitudes are
four-dimensional cut-constructible \refs{\BernZX,\BernCG}.
This means that they
can be completely determined by
studying their finite unitarity cuts, and therefore the dimensional regularization parameter can be set to zero.
  Furthermore, they can be expressed as a linear
combination of known integrals called scalar box, triangle, and
bubble integrals, with rational coefficients in the kinematical
invariants.

The scalar part is more complicated, as only part of it can be
determined by studying four-dimensional unitarity cuts. There are
single-valued pieces that have to be determined using some other
method.\foot{In principle the whole scalar part can be computed
from unitarity cuts if higher orders in $\epsilon$, the
dimensional regularization parameter, are kept. (See section 4.4
of \BernJE.)} The current state of the art in QCD is the
five-gluon amplitude \bernfive.
This means that even the scalar
part has been fully computed for all helicity configurations.

For special helicity configurations much more is known.
$\CA^{\N=1}$ is known for all MHV amplitudes \BernCG. Also in
\BernCG, the cut constructible part of $\CA^{\rm scalar}$ was
given for MHV amplitudes where the gluons of negative helicity are adjacent.
More
recently, the non-adjacent case was computed in \BedfordNH\ using
the techniques of \refs{\WittenNN, \CachazoKJ, \loopmhv}.
Also recently, $\CA^{\N=1}(1^-,2^-,3^-,4^+,5^+,6^+)$ was presented
in \BidderTX\ and then extended to
$\CA^{\N=1}(1^-,2^-,3^-,4^+,\ldots ,n^+)$ in \BidderRI. Also in
\BidderRI, the scalar box coefficients of all other next-to-MHV
six-gluon amplitudes were computed using quadruple cuts \BrittoNC.

For $\CA^{\rm scalar}$, apart from the five-gluon case, all
amplitudes with at most one negative helicity gluon are also known
\refs{\bernplusa,\bernplusb,\MahlonSI}.

In this paper, we introduce a systematic approach to computing any
finite unitarity cut of amplitudes of gluons.
Although our focus is on $\CA^{\N =1}$, it is important to mention
that this can also be applied to obtain the four-dimensional cut
constructible part of $\CA^{\rm scalar}$ or even as an alternative
way of computing $\CA^{\N=4}$ amplitudes.
The basic idea is to
exploit the representation of the Lorentz invariant measure of a
null vector $\ell$, introduced in \CachazoKJ, as a measure over
$\QR^+\times \Bbb{CP}^1\times \Bbb{CP}^1$ with contour of
integration a certain diagonal $\Bbb{CP}^1$. More explicitly, one
writes $\ell_{a\dot a} = t\lambda_a\tilde\lambda_{\dot
a}$, and then
\eqn\expli{ \int d^4\ell \delta^{(+)}(\ell^2) ( \bullet ) =
\int_0^{\infty }t\; dt \int_{\tilde\lambda = \bar\lambda}
\vev{\la, d\la }[\lt ,d \lt ] (\bullet )}
where $(\bullet )$ represents a generic integrand.

 It turns out that the integration on the right hand side of
\expli\ can always be reduced in a systematic way to an integral
performed in \CachazoKJ. The main simplification arises because
the final integrals always localize to some poles in the region of
integration. This is reminiscent of the technique developed in
\CachazoDR\ where certain differential operators are applied to
the cut integral in order to produce a localization via a
holomorphic anomaly \CachazoBY. Surprisingly, here we find that up
to an integration over a single Feynman parameter, which is
responsible for logarithms, all unitarity cuts localize by
themselves without the need of a differential operator.

In the case of unitarity cuts of $\N=1$ amplitudes of gluons,
$\CA^{\N=1}$, one expects bubble, triangle and box scalar
integrals to contribute to a given cut. Remarkably, our procedure
naturally leads to a clean separation of the three kinds of contributions,
allowing for an individual calculation of the corresponding
coefficients.

An important simplification in $\CA^{\N=1}$ is that one- and
two-mass triangle coefficients never need to be computed. It turns
out that their contributions always cancel against singular pieces
in box integrals. This leaves us with bubbles, three-mass
triangles and finite boxes, which we define in detail. Since the
coefficient of scalar boxes can easily be computed from quadruple
cuts \BrittoNC\ one can disregard that piece and concentrate on the
bubble and three-mass triangle scalar integral coefficients.

As an application of our technique we compute all $\CA^{\N=1}$
next-to-MHV six-gluon amplitudes.  The reason we have undertaken
the whole calculation is because six-gluon amplitudes in QCD are
going to be important for future colliders and our computation
completes the second piece in \themiste. It is important to
mention that this calculation requires the use of tree-level
amplitudes of gluons with two fermions or two scalars. Luckily,
very compact formulas for those amplitudes were derived very
recently \refs{\LuoRX,\LuoMY} by extending the techniques of
\refs{\BrittoAP, \BrittoFQ}.

This paper is organized as follows: In section 2, we define
$\CA^{\N=1}$ in terms of a linear combination of scalar integrals.
Using its singular behavior we show that only bubble, three-mass
triangle and finite boxes are necessary. In section 3, we study
general unitarity cut integrals and present the method for
computing them explicitly. For $\CA^{N=1}$ we explain the way to
basically read off the coefficients of bubbles and three-mass
triangles. In section 4, we present the calculation of
$\CA^{\N=1}$ non-MHV six-gluon amplitudes. We also present our results for the next-to-MHV $n$-gluon amplitude where all negative-helicity gluons are consecutive.  This amplitude has appeared in \BidderRI, but by our procedure it emerges in a different form.  Explicit details are
given in order to illustrate the steps described in section 3. In
section 5, we summarize the results for all the amplitudes
computed in section 4. Section 5 is intended to be self contained
so that the reader interested only in the six-gluon amplitude
results can skip the rest of the paper. Appendix A contains a
detailed definition of scalar integrals as well as our definition
of finite box integrals. Appendix B summarizes the results of
tree-level amplitudes of gluons and fermions needed in section 4.


Throughout the paper, we use the following notation and
conventions along with those of \WittenNN\ and the spinor
helicity-formalism \refs{\berends,\xu,\gunion}. The external gluon
labeled by $i$ carries momentum $p_i$. Since $p_i^2=0$, it can be
written as a bispinor $(p_i)_{a\dot a} =
\lambda_{i~a}\tilde\lambda_{i~\dot a}$. Inner product of null
vectors $p_{a\dot a}=\la_a \lt_{\dot a}$ and $q_{a\dot a}=\la'_a
\lt'_{\dot a}$ can be written as $2p\cdot q = \vev{\la , \la'}[\lt
,\lt' ]$, where $\vev{\la , \la' } = \epsilon_{ab}\la^a\la'^b$ and
$[\lt , \lt' ] = \epsilon_{\dot a\dot b}\lt^{\dot a}\lt'^{\dot
b}$. Other useful definitions are:
\eqn\ournot{ \eqalign{ P_{i \ldots j} & \equiv
p_i+p_{i+1}+\cdots+p_j\ \cr K_i^{[r]} &\equiv
p_i+p_{i+1}+\cdots+p_{i+r-1}\ \cr t_i^{[r]} &\equiv
(p_i+p_{i+1}+\cdots+p_{i+r-1})^2\ \cr \gb{i|\sum_r  p_r |j}
&\equiv \sum_r \vev{i~r}[r~j] \cr \vev{i|(\sum_r p_r )(\sum_s p_s
)|j} &\equiv \sum_r\sum_s \vev{i~r}[r~s]\vev{s~j} \cr [i|(\sum_r
p_r )(\sum_s p_s )|j] &\equiv \sum_r\sum_s [i~r]\vev{r~s}[s~j] \cr
\gb{i|(\sum_r p_r)(\sum_s p_s )(\sum_t p_t )|j} &\equiv \sum_r
\sum_s \sum_t \vev{i~r}[r~s]\vev{s~t}[t~j] }}
where addition of indices is always done modulo $n$.


\newsec{One-Loop $\N=1$ Amplitudes}

Amplitudes of gluons at one-loop admit a color decomposition
\refs{\colorord,\mangparke} with single and double trace
contributions. The piece proportional to the single trace term
$\Tr (T^{a_1}\ldots T^{a_n})$ is called the leading color partial
amplitude and it is denoted by $A_{n;1}(1,\ldots ,n)$. In this
paper we concentrate on $A_{n;1}(1, \ldots ,n)$. The reason is
that when all particles in the loop are in the adjoint
representation, all sub-leading color amplitudes are given as
linear combinations of $A_{n;1}$ with permutations of the gluon
labels (See section 7 of \BernZX\ for a proof.) This is the case
for all amplitudes we consider. In the remainder of the paper we
will simplify the subscript and just denote the leading color
partial amplitude by $A_n(1,\ldots, n)$.

We consider amplitudes of gluons where an $\N=1$ chiral multiplet
circulates in the loop. Reduction techniques allow us to express
these amplitudes in terms of scalar integrals in the shapes of
boxes $I_4$, triangles $I_3$, and bubbles $I_2$ \refs{\BernKR,
\BernCG}. These functions are given explicitly in appendix A,
along with some helpful figures.

The amplitude thus takes the following form:
\eqn\ampl{\eqalign{ A^{\N=1}_n = {r_\Gamma (\mu^2)^{\epsilon} \over
 (4\pi)^{2-\epsilon}}\sum\left(c_4^{1m} I_4^{1m} +
c_4^{2m~e} I_4^{2m~e}+ c_4^{2m~h} I_4^{2m~h} + c_4^{3m} I_4^{3m} +
c_4^{4m} I_4^{4m}\right. \cr \left. + c_3^{1m} I_3^{1m}+ c_3^{2m}
I_3^{2m}+ c_3^{3m} I_3^{3m} + c_2 I_2 \right).}}
Here $\epsilon=(4-D)/2$ is the dimensional regularization
parameter, $\mu$ is the renormalization scale, and $r_\Gamma$ is defined by
\eqn\rgamma{r_\Gamma = {\Gamma(1+\epsilon)\Gamma^2(1-\epsilon) \over \Gamma(1-2\epsilon).}
}
The sum runs over all the cyclic permutations within each type of
integral. The coefficients $c$ of the scalar integrals are
rational functions of spinor products. This follows from the
reduction procedure \BernCG.


\subsec{Singular Behavior}

The infrared and ultraviolet singular behavior of these
amplitudes is known and was given in \refs{\kuni,\giel,\BernIX}.
For the case of a gluon amplitude with the $\N=1$ chiral multiplet
in the adjoint representation circulating in the loop, the
divergent behavior is given simply in terms of the tree-level
amplitude by
\eqn\iranduv{ A_{n}^{\N=1}|_{\rm singular} = {r_\Gamma \over
\epsilon (4\pi)^{2-\epsilon}} A_{n}^{\rm tree}, }
where
\eqn\rgamma{r_\Gamma = {\Gamma(1+\epsilon)\Gamma^2(1-\epsilon) \over \Gamma(1-2\epsilon)}
}
and $A_{n}^{\rm tree}$ is the color-ordered tree-level amplitude.

Now let us see what this means for the scalar integral
coefficients.  The integrals $I_3^{3m}$ and $I_4^{4m}$ are finite.
Therefore their coefficients do not contribute to \iranduv. The
bubble integral diverges as $1/\epsilon$. One- and two-mass
triangle integrals can be conveniently written in terms of the
function
\eqn\bubb{ T(s) = {r_\Gamma\over \epsilon^2}(-s)^{-\epsilon}}
as follows:
\eqn\onet{ I^{1m}_{3}(s) = {1\over (-s)}T(s), \qquad I^{2m}_{3}(s,t)
= {1\over (-s)-(-t)}\left( T(s) - T(t) \right)}
where $s$ and $t$ denote the invariants in different independent
channels.

Finally, one-, two-, and three-mass box scalar integrals have the
property that they can be made finite by adding linear
combinations of $T(s)$ functions. We denote the finite box
integral functions by $I_{4F}$, where $F$ stands for finite. Their
definition is given in detail in appendix A.

Now we are ready to derive the main result of this section. From
\iranduv\ we see that all divergences of the form $1/\epsilon^2$
must be absent. This implies that all $T(s)$ functions must cancel
among the different terms. Since one- and two-mass triangle
integrals are given entirely as linear combination of $T(s)$
functions with rational coefficients, it follows that their
coefficients are such that they do not appear in the final answer
for the amplitude. Our interpretation is that the only reason they
must be included is to cancel the $1/\epsilon^2$ divergences from
the box integral.

Therefore, we reach the conclusion that $\CA_{n}^{\N=1}$ can be
written as a linear combination of finite box scalar integrals
$I_{4F}$, three-mass triangles $I^{3m}_3$ and bubbles $I_2$. More
explicitly,
\eqn\kole{ \CA_{n}^{\N=1} = {r_\Gamma (\mu^2)^{\epsilon} \over
 (4\pi)^{2-\epsilon}} \sum\left( c_2 I_2 + c_3^{3m} I_3^{3m}
+ c_4^{1m} I_{4F}^{1m} + c_4^{2m~e} I_{4F}^{2m~e}+ c_4^{2m~h}
I_{4F}^{2m~h} + c_4^{3m} I_{4F}^{3m} + c_4^{4m} I_{4}^{4m}
\right).}

Finally, among the bubble coefficients there is one relation that
must hold in order to satisfy \iranduv, namely that the sum of all bubble coefficients reproduces the tree-level amplitude:
\eqn\irrel{\sum c_{2} =A_{n}^{\rm tree}.}
In section 4, \irrel\ is used as a very non-trivial
consistency check of our results for next-to-MHV six gluon
amplitudes.

\newsec{Coefficients from Unitarity Cuts}

In this section we introduce a new method for computing explicitly
any finite unitarity cut in a gauge theory with massless particles
running in the loop. Of course, our aim here is to apply the
technique to the computation of the coefficients in \kole\ which
determine $\CA^{\N =1}$. Nevertheless, it is important to mention
that this can also be applied to obtain the four-dimensional cut
constructible part of $\CA^{\rm scalar}$ or even as an alternative
way of computing $\CA^{\N=4}$ amplitudes.

 The unitarity
cut in the $(i,i+1,\ldots , j-1,j)$-channel is computed by cutting
two propagators in the loop whose momenta differ by
$P_{ij}=p_i+\ldots +p_j$ in all Feynman diagrams contributing to
the amplitude. Adding up all these contribution we find a ``cut
integral" \cuts\foot{Further information about this technique may be found in
\sbook.
This body of work was not intended to apply to massless theories.
  We find that the material can nevertheless be adapted
for the considerations of this paper.  The modern interpretation is found in
\refs{\BernZX,\BernCG}.}

\ifig\convi{Representation of the cut integral. Left and right
tree-level amplitudes are on-shell. Internal lines represent the
legs coming from the cut propagators.}
{\epsfxsize=0.50\hsize\epsfbox{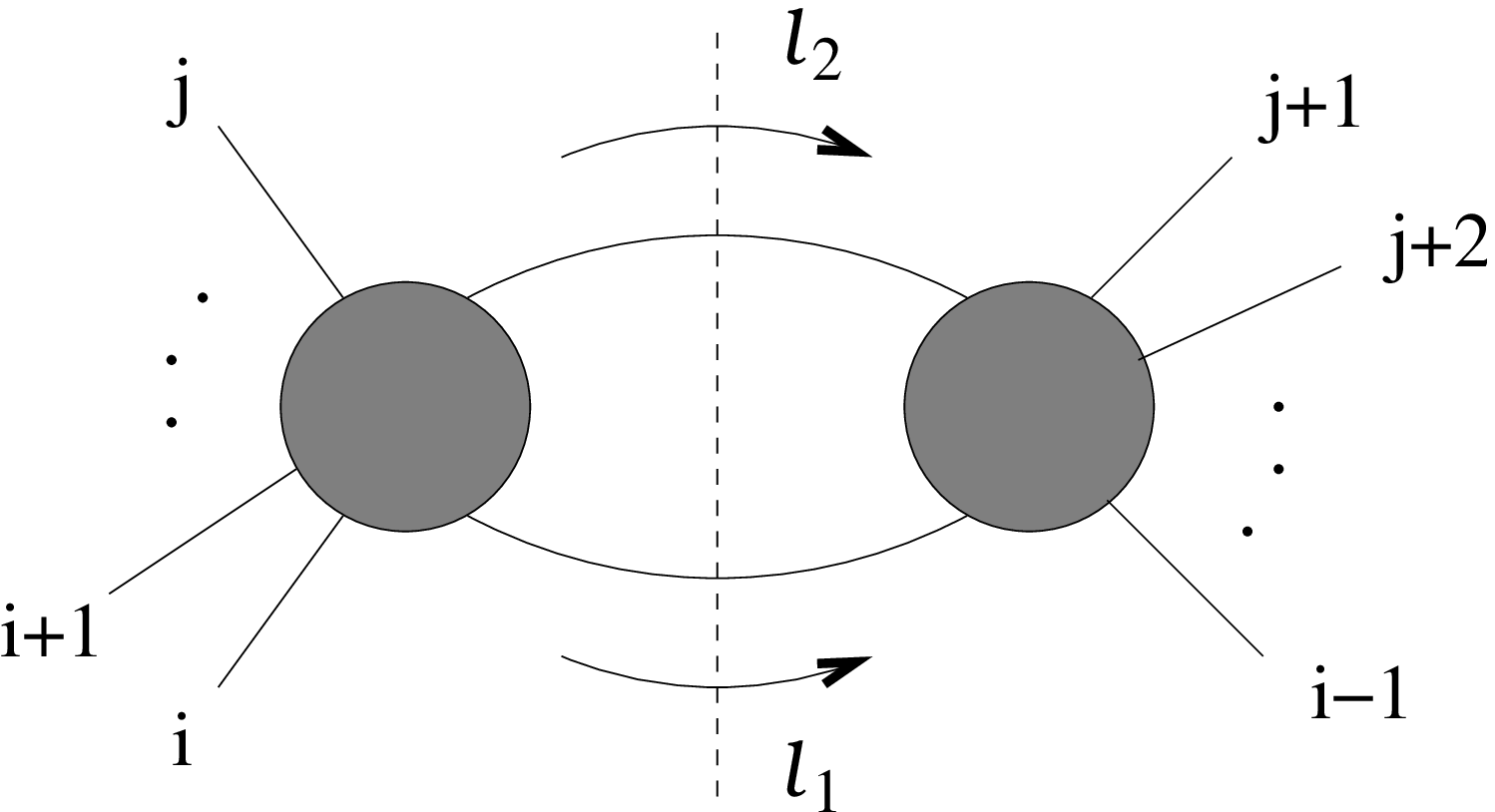}}

\eqn\cutIn{ \eqalign{ & C_{i,i+1,\ldots ,j-1,j} = \cr &   \int
d\mu A^{\rm tree}(\ell_1,i,i+1,\ldots ,j-1,j,\ell_2)A^{\rm
tree}((-\ell_2),j+1,j+2,\ldots ,i-2,i-1,(-\ell_1)),}}
where $d\mu= d^4\ell_1 d^4\ell_2
\delta^{(+)}(\ell_1^2)\delta^{(+)}(\ell_2^2)\delta (\ell_1+\ell_2
- P_{ij})$ is the Lorentz invariant phase space measure of two
light-like vectors $(\ell_1, \ell_2)$ constrained by momentum
conservation.  See \convi.

This cut integral computes the discontinuity of the amplitude
across a given branch cut in the space of kinematical invariants.
The same discontinuity can be computed from the right hand side of
\ampl. Note that the coefficients are rational and thus do not
have branch cuts. The scalar integrals have a cut in this channel
only if they contain the same two propagators that are cut.

The idea is to determine the cuts of the known scalar integrals,
compute \cutIn\ explicitly and then solve for the coefficients by
comparing both sides.

The phase space integral in \cutIn\ can be performed following the
techniques of \refs{\CachazoKJ}. The idea is to use momentum
conservation to write the integral entirely in terms of just one
of the cut propagator momenta, say $\ell=\ell_1$.  This vector is
then parametrized as $\ell_{a\dot a}=t\la_a\lt_{\dot a}$, where
the scale $t$ is real and the spinors $\la$ and  $\lt$ are
independent homogeneous coordinates on two copies of $\Bbb{CP}^1$.
The integral is then performed over the diagonal $\Bbb{CP}^1$
defined by $\tilde\lambda =\bar\lambda$. The integral can be
rewritten as
\eqn\meas{ \int d^4\ell \delta^{(+)}(\ell^2) ~ (\bullet ) =
\int_0^{\infty}dt~t\int\vev{\lambda,
d\lambda}[\tilde\lambda,d\tilde\lambda] ( \bullet ),}
where the bullets represent generic arguments.

We first illustrate the procedure by computing the cuts of bubble
and three-mass triangle scalar integrals entering in \kole\ and
then we discuss the calculation of the general cut integral
\cutIn.

\subsec{Bubble and Three-Mass Triangle Unitarity Cuts}

As a warm-up let us compute the double cut of a bubble and a
three-mass triangle using \meas.  We will need these results
to read off the coefficients we need.   Let us denote the cut of a scalar
integral $I$ by $\Delta I$.

The unitarity cut of a bubble is given by (see appendix A for a
definition of the original integral\foot{In the following calculations, we are omitting the factor of $-i {(4\pi)^{2-\epsilon} \over (2\pi)^{4-2\epsilon}}$ which has been accounted for in the overall factor in \ampl\ and \kole.})
\eqn\bubint{\eqalign{ \Delta I_{2}(K) & = \int d^4 \ell
\delta^{(+)}(\ell^2) \delta^{(+)}((\ell-K)^2) \cr & =
\int_0^\infty t dt \int \braket{\la~d\la}[\W \la~d\W \la]
\delta^{(+)}( K^2- t K_{a \D a}\la^a \W\la^{\D a}) \cr & =  \int
\braket{\la~d\la}[\W \la~d\W \la] {K^2\over (K_{a \D a}\la^a
\W\la^{\D a})^2}}} were we have used that in the kinematic regime
where $K^2>0$ the delta function always has its support in the
integration region of $t$.

We postpone evaluating the last integral in \bubint.  First let us
evaluate a slightly more complicated integral \CachazoKJ\foot{This
integral was performed as part of a derivation of MHV diagrams for
tree-level amplitudes of gluons from a twistor string theory
calculation \WittenNN.},
\eqn\basic{{\cal I} = \int \braket{\la~d\la}[\W \la~d\W \la]
{1\over (K_{a \D a}\la^a \W\la^{\D a})^2}g(\lambda ) }
with
\eqn\gfor{ g(\lambda ) = {\prod_{i=1}^k \vev{\lambda , A_i}\over
\prod_{j=1}^k \vev{\lambda , B_j }}}

First note the following identity that holds for an arbitrary but
fixed negative chirality spinor $\eta$:
\eqn\ideo{{[\W \la~d\W \la]\over (K_{a \D a}\la^a \W\la^{\D
a})^2}g(\lambda ) = -d\lt^{\dot c}{\del\over \del\lt^{\dot
c}}\left( {[\lt ,\eta ]\over (K_{a\dot a}\la^a\lt^{\dot a})
(K_{a\dot a}\la^a\eta^{\dot a})} g(\lambda ) \right). }
This identity holds for all values of $\lambda$ except for those
where the denominator vanishes along the contour of integration.
The reason is that along the contour of integration $\lt =
\bar\lambda$ and therefore
\eqn\holo{-d\lt^{\dot c}{\del\over \del\lt^{\dot c}}{1\over
\vev{\lambda , \zeta }} = 2\pi \bar\delta (\vev{\lambda , \zeta
}), }
where we have introduced a $(0,1)$-form $\bar\delta(\vev{\lambda , \zeta
}),$ such that
\eqn\defbardel{
\int \vev{\lambda~d\lambda} ~\bar\delta(\vev{\lambda , \zeta
}) B(\lambda) = -i B(\zeta).
}

Let us write the complete form of \ideo\ that is valid for all
values of $\lambda$ along the contour of integration:
\eqn\valid{ \eqalign{{[\W \la~d\W \la]\over (K_{a \D a}\la^a
\W\la^{\D a})^2}g(\lambda ) = & -d\lt^{\dot c}{\del\over
\del\lt^{\dot c}}\left( {[\lt ,\eta ]\over (K_{a\dot
a}\la^a\lt^{\dot a}) (K_{a\dot a}\la^a\eta^{\dot a})} g(\lambda )
\right) \cr & + {2\pi [\lt, \eta]\over K_{a\dot a}\la^a\lt^{\dot
a}}\left( -\bar\delta (K_{a\dot a}\la^a\eta^{\dot a})g(\la )+
{1\over K_{a\dot a}\la^a\eta^{\dot a}}\sum_{j=1}^k
\bar\delta(\vev{\la , B_j})g(\lambda)\vev{\la ,B_j}\right). } }
The contribution from the first term in \valid\ gives zero after
integration over $\lambda$. On the other hand, the delta functions
in the remaining terms localize the $\lambda$ integral and give
the value of the integrand at the pole. More explicitly one uses
that
\eqn\usal{ \int \vev{\la ,d\la} \bar\delta(\vev{\la, \lambda_B})
H(\la ) = - i H(\lambda_B). }

A short calculation reveals that \basic\ is given by
\eqn\fin{ {\cal I} = -{1\over K^2}g(\lambda_K) + \sum_{j=1}^k
{[B_j,\eta]\over
\gb{B_j|K|B_j}\gb{B_j|K|\eta}}{\prod_{i=1}^k\vev{B_j,A_i}\over
\prod_{l\neq j}\vev{B_j, B_l}}}
where $\lambda_{K\; a} =K_{a\dot a}\eta^{\dot a}$.

This is the basic result that will allow us to calculate any
double cut in section 3.2.

Going back to the cut of the bubble integral \bubint\ we find that
by setting $g(\lambda) =1$ in \fin\
\eqn\goal{ \Delta I_2(K) = -1.}

Consider now a three-mass triangle integral.
\ifig\upionc{A double cut of a three-mass triangle integral.}
{\epsfxsize=0.20\hsize\epsfbox{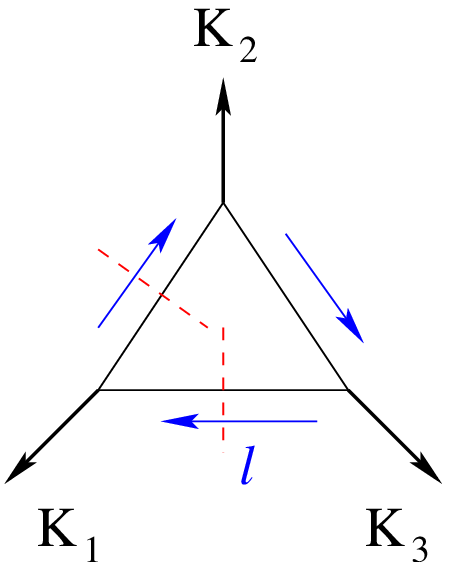}}
 Denote the momenta at
the vertices by $K_1$, $K_2$ and $K_3$. Let us calculate the cut
in the $K_1$ channel, $\Delta_1$. Let the momentum in the cut
propagators be $\ell$ and $\ell-K_1$.  See \upionc.  Then
\eqn\triint{\eqalign{ \Delta_1 I^{3m}_{3} &=- \int d^4 \ell
\delta^{(+)}(\ell^2) {\delta^{(+)}((\ell-K_1)^2) \over
(\ell+K_3)^2} \cr & = -\int_0^\infty t dt \int \braket{\la~d\la}[\W
\la~d\W \la] {\delta( K_1^2- t K_{1,a \D a}\la^a \W\la^{\D a})
\over K_3^2+ t K_{3,a \D a}\la^a \W\la^{\D a}}  \cr & =-  \int
\braket{\la~d\la}[\W \la~d\W \la] {K_1^2\over (K_{1,a \D a}\la^a
\W\la^{\D a})^2} { (K_{1,a \D a}\la^a \W\la^{\D a})\over K_3^2
(K_{1,a \D a}\la^a \W\la^{\D a}) + K_1^2 (K_{3,a \D a}\la^a
\W\la^{\D a})} \cr & =  - \int \braket{\la~d\la}[\W \la~d\W \la]
{1\over (K_{1,a \D a}\la^a \W\la^{\D a}) (Q_{a \D a}\la^a
\W\la^{\D a})} }} where $Q_{a \D a}= {K_3^2\over K_1^2}(K_{1,a \D
a})+ (K_{3,a \D a})$.

This integral can be brought to the same form as that in \bubint\
by introducing a Feynman parameter to combine the two denominators
into one. This Feynman parameter integration turns out to be the
only one needed in the calculation of general cut integrals in
section 3.2. This is why we show it explicitly here.
\eqn\qeqo{{1\over (K_{1,a \D a}\la^a \W\la^{\D a}) (Q_{a \D
a}\la^a \W\la^{\D a})} = \int_{0}^1 dx {1\over (((1-x) K + x Q )_{a
\D a}\la^a \lt^{\dot a} )^2 } }
Performing first the $\lambda$ and $\lt$ integrations, we get
\eqn\mmbpo{\int_{0}^1 dx {1\over ((1-x) K + x Q)^2}.
}
The result of the $x$ integration in the kinematic regime corresponding to the
channel under consideration, i.e., where $K_1^2>0$ and both
$K_2^2<0, K_3^2<0$ is then:
\eqn\moretriint{\Delta_1 I_3^{3m} =  {1\over \sqrt{\Delta}}
\left(\ln { 2 Q^2 -\sqrt{\Delta} \over 2Q^2 +\sqrt{\Delta}}-\ln{ 2
( -K_2 \cdot K_3-K_1^2)-\sqrt{\Delta} \over   2 (-K_2 \cdot K_3
-K_1^2)+\sqrt{\Delta}} \right), } where $\Delta$ denotes the
discriminant that arises from the quadratic equation in the
Feynman parameter,  given by
\eqn\discrim{ \Delta  = (K_1^2)^2 + (K_2^2)^2+(K_3^2)^2 - 2
K_1^2K_2^2- 2 K_2^2K_3^2 - 2 K_3^2 K_1^2 . }
%

\subsec{General Cut Integrals}

As discussed in section 2, calculating all one-loop $\N=1$
amplitudes is equivalent to finding the rational coefficients of
boxes, three-mass triangles and bubbles. The box coefficients can
be computed by quadruple cuts. In principle, the three-mass
triangle coefficients can be computed using triple cuts and the
bubble coefficients require double cuts. In practice we find that
all bubble and three-mass triangle coefficients can be computed
from double cuts in a simple and systematic manner. It turns out
that the separation of the three-mass and bubble integral
coefficients is easily done because of the striking difference in
the form of their cuts; see \goal\ and \moretriint.

In the remainder of this section, we explain how to perform
general double cut integrals \cutIn.

First, find spinor-product expressions for the two tree-level
amplitudes that form the integrand of \cutIn. These are tree-level
amplitudes of gluons with two fermions or two scalars. Two recent
techniques allow the calculation of those amplitudes: MHV diagrams
\refs{\GeorgiouWU, \WuFB, \WuJX, \GeorgiouBY} and recursion
relations \refs{\LuoRX, \LuoMY}. However, at this point all we
need is that they are rational functions in the spinor products.

Recall that the measure in \cutIn\ is $d\mu = d^4\ell_1 d^4\ell_2
\delta^{(+)}(\ell_1)\delta^{(+)}(\ell_2)\delta^{(4)}(\ell_1+\ell_2-P_{ij})$,
where $P_{ij} = p_i+p_{i+1}+\ldots + p_j$. Using the last delta
function to perform the $\ell_2$ integration and \meas\ to write
the measure over $\ell_1$, we find
\eqn\hala{C_{i,\ldots ,j} = \int_0^{\infty}t dt \vev{\lambda ,
d\lambda } [\tilde\lambda , d\tilde\lambda ] \delta^{(+)}( t
\lambda_a\tilde\lambda_{\dot a}P_{ij}^{a\dot a} - P_{ij}^2)
G(\lambda , \tilde\lambda , t).}
We denote $\ell_1$ by $\ell$ when there is no possibility of
confusion. Recall that $\ell_{a\dot a} =t\la_a \lt_{\dot a}$.
$G(\lambda , \tilde\lambda , t)$ is the function that arises from
the product of the two tree-level amplitudes in \cutIn. A simple
observation that helps in actual calculations is that in order to
obtain $G(\lambda , \tilde\lambda , t)$, one has to write
expressions of the form $\vev{\bullet , \ell_2}$ or $ [\bullet ,
\ell_2]$ in terms of $\ell = \ell_1$. A systematic way of doing
this is by using the following identity:
\eqn\syst{ \vev{\bullet , \ell_2} = {\vev{\bullet , \ell_2}
[\ell_2 , \ell_1]\over [\ell_2 , \ell_1]} = {\gb{\bullet | \ell_2
 | \ell_1}\over [\ell_2 , \ell_1]} = {\gb{\bullet | P_{ij}|\ell_1}\over [\ell_2 , \ell_1]}.}
A similar identity is valid for $[\bullet , \ell_2]$. The factors
$[\ell_2 , \ell_1]$ and $\vev{\ell_2 , \ell_1}$ all pair up in the
end allowing for the use of the vector form of $\ell_2$. This
happens because the product of the amplitudes must be invariant
under the scaling $z\lambda_{\ell_2}$ and
$z^{-1}\tilde\lambda_{\ell_2}$.

Going back to the integral \hala, let us perform the $t$
integration by using the delta function,
\eqn\qeli{ C_{i,\ldots ,j} = P^2_{ij} \int { \vev{\lambda ,
d\lambda } [\tilde\lambda , d\tilde\lambda ] \over (P_{ij}^{a\dot
a}\lambda_a\tilde\lambda_a)^2 } G \left(\lambda , \tilde\lambda ,
{ P^2_{ij}\over P_{ij}^{a\dot a}\lambda_a\tilde\lambda_a }\right)
.}

Let us assume that the $\tilde\lambda$ dependence in the
denominator of $G$ is simpler, i.e. it has fewer factors than that
of $\lambda$. If the opposite were true, we would use the conjugate
of the discussion that follows.

Since $\lambda$ and $\lt$ are independent homogeneous coordinates
on two $\Bbb{CP}^1$ one must require $t$ to transform as $t\to
(wz)^{-1}t$ when $(\la, \lt)\to (w\la, z\lt)$ so that $\ell_{a\dot
a} = t\la_a\lt_{\dot a}$ remains invariant. For the integral
\qeli\ to make sense, it must be the case that $G(\lambda ,
\tilde\lambda , t)$ is invariant under the scaling
$(\tilde\lambda, t) \to ( z\tilde\lambda, z^{-1}t)$. This ensures
that $G$ in \qeli\ has degree zero in $\tilde\lambda$. This
implies that it can be written as a sum of terms of the form
\eqn\term{ {\prod_l [A_l, \ell]\over \prod_i \gb{\ell | Q_i | \ell
}\prod_j [A_j, \ell]} g(\lambda ). }
Each term has degree zero in $\tilde\lambda$. The function
$g(\lambda)$ contains all other terms that do not depend on $\lt$.

There are two ways to proceed at this point. One is based on the
introduction of several Feynman integration parameters. We find
that this method becomes very cumbersome as the number of Feynman
parameters increases. The second approach keeps the number of
Feynman integrations to be at most one, and it has the advantage of
leading to a clean separation of bubble, three-mass triangle and
box coefficients. We discuss both approaches because they might be
useful in different situations.

\subsubsec{Feynman Parametrizations}

We want to transform the denominator of \term\ by replacing every
factor $1/[A_i, \ell ]$ by $-\vev{A_i ,\ell }/ \gb{\ell | A_i |
\ell}$ and then using Feynman parameters to combine all factors in the
denominator, including the one in \qeli, into one of the form
$1/\gb{\ell | T(x_1,\ldots ,x_{m+2})|\ell}^{m+2}$.

The integral to be performed has now the form
\eqn\newf{\int \prod_{i=1}^{m+2} dx_i \delta \left(
\sum_{j=1}^{m+2} x_j -1 \right)\int { \vev{\lambda , d\lambda }
[\tilde\lambda , d\tilde\lambda ] \over \gb{\ell | T(x_1,\ldots
,x_{m+2})|\ell}^{m+2}}\prod_{l=1}^m [A_l, \ell] \tilde g(\lambda ) }
where we have absorbed all $\vev{A_i,\ell }$ into $\tilde
g(\lambda)$. Recall that the total degree in $\lt$ of the
integrand must be $-2$. Therefore, there are $m$ factors in the
numerator containing $\lt$.

It is not difficult to find an analog of \ideo, i.e., to write the
integrand as a total derivative in $\lt$. Indeed, one can prove by
induction that
\eqn\howto{\eqalign{ &  { [\lt~d\lt] \prod_{i=1}^j [A_i~\ell]
[\eta ~\ell]^{m-j} \over \gb{\ell|T|\ell}^{m+2}}  \cr & =
[d\ell~\partial_\ell] \left[ {\prod_{i=1}^{j} \gb{\ell|T| A_i}
\over \gb{\ell|T|\ell}^{m+1}} \left( \sum_{k=0}^j { (-1)^{j-k}
(j-k)!\over (m+1-j)...(m+1-k)} g_k[x_i] {[\eta ~\ell]^{m+1-k}
\over \gb{\ell|T|\eta }^{j+1-k}}\right) \right]. }}
Here, $\eta$ is an arbitrary but fixed spinor and
\eqn\kiki{ g_k[x_s] = \sum_{i_1<i_2 < \ldots <i_k}x_{i_1}\ldots
x_{i_k} \quad {\rm with} \quad x_i = {[A_i, \ell ]\over \gb{\ell |
T | A_i}}.  }

Now we can proceed as explained in section 3.1. The idea is to
realize that \howto\ is valid for all values of $\la$ except those
for which there is a pole. The final value of the integral comes
entirely from those poles as in \valid\ that led to \fin.

The complication with this approach is that the final result is
expressed in terms of several Feynman parameter integrations. In
practice, if the number of Feynman parameters is two or more then
the integration procedure is cumbersome.

\subsubsec{Simple Pole Expansion}

In order to avoid the proliferation of Feynman parameters we
propose a second way of treating the integral \qeli. The idea is
that before doing the integral, we should separate the denominator
factors with $\lt$ as much as possible, at the cost of more terms.
Where there is a product $[a~\ell][b~\ell]$ in the denominator,
multiply both numerator and denominator by $[a~b]$. Then apply
Schouten's identity, \eqn\schou{[i~j][k~l]=[i~k][j~l]+[i~l][k~j],}
in the numerator with another factor $[c~\ell]$ (which must exist
by homogeneity). We then get two terms with $[a~\ell]$ or
$[b~\ell]$ in the numerator, cancelling one of the denominator
factors.  The result, in terms of $\lt_{\ell}$, is a denominator
of the form $\prod_r\gb{\ell|Q_r|\ell} [A ~\ell]$ in every term.

Factors of the form $\gb{\ell|Q_r|\ell}$ can be treated in the
same way by writing
\eqn\lasi{ \gb{\ell|Q_r|\ell} = [\tilde Q_r ,\ell ]}
where $\tilde\lambda_{Q_r \;\dot a} = -(Q_{r})_{a\dot
a}\lambda^a_\ell$.

Using this procedure to split poles in $\lt$ and given that the
integrand has degree $-2$ in $\lt$, one finds in the end only two
possible kind of integrals
\eqn\kindt{{\cal I}_A = \int { \vev{\lambda , d\lambda }
[\tilde\lambda , d\tilde\lambda ] \over (P_{ij}^{a\dot
a}\lambda_a\tilde\lambda_a)^2 }H(\lambda);\qquad  {\cal I}_B =
\int { \vev{\lambda , d\lambda } [\tilde\lambda , d\tilde\lambda ]
\over (P_{ij}^{a\dot a}\lambda_a\tilde\lambda_a)^2 }{[B , \ell
]\over \gb{\ell | Q_r | \ell } }H(\lambda) . }

For the second class of integrals we can apply the splitting
procedure once more to split the product $\gb{\ell | P |\ell
}\gb{\ell | Q_r | \ell }$. Then we find an integral of the form
$\CI_A$ and one of the form
\eqn\newi{ {\cal I}_C = \int { \vev{\lambda , d\lambda }
[\tilde\lambda , d\tilde\lambda ] \over \gb{\ell | P |
\ell}\gb{\ell | Q_r | \ell } }H(\lambda).}
Here $H(\lambda)$ is used to denote a generic function of $\ell$
and independent of $\tilde\lambda$.

As anticipated in section 3.1, the two kind of integrals, i.e.,
${\cal I}_A$ and ${\cal I}_C$, appeared in the calculation of the
bubble and three-mass triangle integral discontinuities. Therefore
one can repeat the same procedure for their computation. In
particular, ${\cal I}_A$ does not require any Feynman parameters
and produces a rational function. On the other hand, ${\cal I}_C$
only requires one Feynman parameter and produces only logarithms.

\subsubsec{Canonical Decomposition}

The decomposition of the cut integral into integrals of the form
${\cal I}_A$ and ${\cal I}_C$ has a very useful byproduct. Since
${\cal I}_A$ produces only rational functions and ${\cal I}_C$
produces only logarithms, it is easy to conclude that the bubble
coefficient is given by
\eqn\bubble{ c_2 = \int { \vev{\lambda , d\lambda } [\tilde\lambda
, d\tilde\lambda ] \over (P_{ij}^{a\dot
a}\lambda_a\tilde\lambda_a)^2 }H(\lambda). }
Here we have used the fact that for a given unitarity cut, there
is only one bubble integral with the corresponding branch cut.

In the calculation of ${\cal I}_C$ one has to introduce a Feynman
parameter $x$ to write the factors with $\tilde\lambda$ in the
denominator as $\gb{ \ell| x P+(1-x)Q | \ell}^2$. The integration
over $\lambda$ and $\tilde\lambda$ produces two different kind of
terms. The first comes from the pole $\gb{ \ell | x P+(1-x)Q |\eta
}$. The second kind comes from the poles in $H(\lambda)$. This is
explained in detail in section 3.1 where \fin\ is computed. There
the two kind of terms are shown explicitly.

Let us write the contribution to $\CI_C$ from the first term in
\fin,
\eqn\jujo{ \int_0^1 dx {1\over (x P+(1-x)Q)^2}H(\lambda(x))
}
where $\lambda(x)$ is the solution of $\gb{ \ell | x P+(1-x)Q
|\eta } = 0 $.

It is easy to see that the contributions to $\CI_C$ from the poles
of $H(\lambda)$ will only have linear factors in $x$ in the
denominator. This implies that upon integration in $x$ they can
only produce logarithms of rational functions of the kinematical
invariants. These simple functions come from the discontinuity of
one-, two-, and three-mass scalar box integrals.

On the other hand, the term given in \jujo\ can produce a more
complicated object. There are again two cases: if the discriminant
of the quadratic equation $(x P+(1-x)Q)^2=0$ is a perfect square,
then we find more one-, two-, and three-mass scalar box
contributions. If the discriminant is not a perfect square then we
either have the contribution of a three-mass triangle or a
four-mass box integral.

Recall that the coefficient of scalar box integrals can be
computed very efficiently by using the quadruple cut technique
introduced in \BrittoNC. This implies that we can ignore all those
contributions and concentrate only on the three-mass triangle
coefficients.

The discriminant of $(x P+(1-x)Q)^2$ is given by
\eqn\diss{ \Delta = 4 ( ((P-Q)\cdot Q)^2 -  (P-Q)^2 Q^2 ).}
Therefore, we are only interested in integrals that produce a
discriminant of the form \discrim, i.e.
\eqn\threem{  \Delta_{3m} = (K_1^2)^2 + (K_2^2)^2+(K_3^2)^2 - 2
K_1^2K_2^2- 2 K_3^2K_1^2 - 2 K_2^2 K_3^2.}

{}From the calculation of the three-mass triangle cut in section
3.1 we know that if $P=K_1$ then in order to produce \threem\ we
need $Q_{a\dot a}={K_3^2\over K_1^2}(K_1)_{a\dot a} + (K_3)_{a\dot
a}$.

Once we have identified the integral that has the discriminant of
a three-mass triangle scalar integral we have to perform one more
decomposition. The idea is to expand the integrand in simple
fractions, as a function of $x$, until we find a term of the form
\eqn\qeqi{c^{3m}_3 \int_0^1 dx {1 \over (x P+(1-x)Q)^2} }
whose coefficient we can identify with the three-mass triangle
coefficient $c^{3m}_3$. This procedure is applied in detail in the
calculation of the cut $C_{23}^{(3;3)}$ in section 4.2.

We illustrate all the features of this procedure for calculating
bubble and three-mass triangle coefficients in the next section
with the example of the six-gluon amplitude. One particularly
challenging technical point is that generically one might expect
poles in the denominator of the form $\vev{\ell |P Q |\ell}$. This
is a quadratic equation for $\la$. We explain how to deal with
this in the calculation of the coefficient $c_{2:2;2}^{(3)}$ in
section 4.2.

\newsec{Example: All $\N=1$ Next-To-MHV Six-Gluon Amplitudes}

There are several pieces missing in the calculation of the
six-gluon next-to-leading order  scattering amplitude in QCD. They are: next-to-MHV
$A^{\N=1}$, MHV and next-to-MHV $A^{\rm scalar}$. Recently
important progress has been made for the first class of
amplitudes. In \BidderTX, the amplitude $A(1^-,2^-,3^-,4^+,5^+,6^+)$ was
presented. In \BidderVX, all scalar box coefficients of the remaining
helicity configurations were computed using quadruple cuts. It
is the aim of this section to present all the remaining
coefficients and some new forms for known ones. These results
complete the next-to-MHV $A^{\N=1}$ piece.

\subsec{First Configuration: $A(1^-,2^-,3^-,4^+,5^+,6^+)$}

This amplitude has been computed in \BidderTX.  Here we rederive
it to illustrate our integration technique.  Our result will agree
with \BidderTX\ but emerge in a slightly different form.

The first observation is that all the box and triangle
coefficients vanish.  This is easily seen by examining the helicity
assignments in all possible distributions.

Thus every double cut simply gives  the coefficient of
the associated bubble integral.  The nonvanishing cuts are
$C_{34}, C_{61}, C_{234}$ and $C_{345}$.  Among these,
cuts $C_{234}$ and $C_{345}$ are mapped to each other by the
permutation of indices
$P_\a:1\leftrightarrow 6,2\leftrightarrow 5,3\leftrightarrow 4 $
plus conjugation, while cuts $C_{34}$ and $C_{61}$ are invariant under $P_\a$.
There is another permutation $P_\b: 1\leftrightarrow 3,4\leftrightarrow 6 $
under which $C_{234}$ and $C_{34}$ are mapped to $C_{345}$
and $C_{61}$ respectively. So there are only two
independent integrands which are  given as
\eqn\ascarid{\eqalign{
C_{34}  = &  \int
d\mu A^{\rm tree}(\ell_1,5,6,1,2,\ell_2)A^{\rm
tree}((-\ell_2),3,4,(-\ell_1)) \cr
= & \int d\mu { \gb{3|1+2|6}^2 \over [6~1][1~2]\braket{3~4}
\gb{5|6+1|2} P_{612}^2} { \gb{\ell_1|1+2|6} [\ell_2~4]\over
\braket{\ell_1~5}[\ell_2~3]} \cr
&  + \int d\mu { \gb{1|5+6|4}^2 \over \braket{5~6}\braket{6~1}[3~4]
\gb{5|6+1|2} P_{561}^2} { \gb{1|5+6|\ell_2} \braket{3~\ell_1}
\over [2~\ell_2] \braket{4~\ell_1}}
}}
and
\eqn\ascaridthree{\eqalign{
C_{612} = & -\int d\mu {  \gb{3|P_{345}|6}^2 \over [6~1][1~2]
\braket{3~4}\braket{4~5} P_{345}^2} { \gb{\ell_1|P_{345}|6}
\braket{3~\ell_1}
\over \gb{\ell_1|P_{345}|2} \braket{5~\ell_1}}
}}
$C_{612}$. When we calculate the integrand, we need to
use the tree-level amplitude of four gluons and a pair of fermions
and complex scalars. These amplitudes are given in
Appendix B.

Now we will do the integration. To demonstrate our method, we
will do one integration (for the cut $C_{34}$) in detail and then simply cite the other three results, which are obtained similarly.

\subsubsec{The Cut $C_{34}$:}

The integrand is given by \ascarid. There are two terms which
are mapped to each other under $P_\a$ so that the cut
$C_{34}$ is invariant. Thus we can focus on the first term only.
In the first term, multiply numerator and denominator by
$\vev{\ell_1~\ell_2}$ and  perform the $t$ integral to get
\eqn\spirurone{\eqalign{
C_{34}^{(1)}  = & -{ \gb{3|P_{12}|6}^2 P_{34}^2\over [6~1][1~2]
\braket{3~4}
\gb{5|P_{61}|2} P_{612}^2} \int \vev{\la_{\ell_1}~d\la_{\ell_1}}
[\la_{\W \ell_1}~d\la_{\W \ell_1}]{ \gb{\ell_1|P_{12}|6} \vev{\ell_1~3} \over \vev{\ell_1~5}
\vev{\ell_1~4}} {1\over \gb{\ell_1|P_{34}|\ell_1}^2} \cr
}}
We can rewrite this as
\eqn\spirurtwo{\eqalign{
C_{34}^{(1)}  = & C \int \vev{\la_{\ell_1}~d\la_{\ell_1}}
[d\W \la_{\ell_1}~\partial_{\W \la_{\ell_1}}]
\left( { \gb{\ell_1|P_{12}|6} \vev{\ell_1~3} \over \vev{\ell_1~5}
\vev{\ell_1~4}} {[\eta~\ell_1]\over \gb{\ell_1|P_{34}|\ell_1}
\gb{\ell_1|P_{34}|\eta}}\right),
}}
where we have defined the constant
 $$C= -{ \gb{3|P_{12}|6}^2 P_{34}^2\over [6~1][1~2]
\braket{3~4}
\gb{5|P_{61}|2} P_{612}^2}.$$
At this stage, we have three poles: $\ket{\ell_1}=\ket{4}$,
$\ket{\ell_1}=\ket{5}$, and $\ket{\ell_1}=|P_{23}|\eta]$. However,
since $|\eta]$ is an arbitrary spinor, we can choose it to be
$|\eta]=|4]$. It is easy to see that after making the above choice we reduce
the integrand into
\eqn\spirurthree{\eqalign{
C_{34}^{(1)}  = & C \int \vev{\la_{\ell_1}~d\la_{\ell_1}}
[d\W \la_{\ell_1}~\partial_{\W \la_{\ell_1}}]
\left( { \gb{\ell_1|P_{12}|6} [4~\ell_1] \over [3~4]\vev{\ell_1~5}
\vev{\ell_1~4}\gb{\ell_1|P_{34}|\ell_1}}\right)
}}
where only one pole $\ket{\ell_1}=\ket{5}$ gives nonzero contribution
(for the pole $\ket{\ell_1}=\ket{4}$, since the factor $ [4~\ell_1]$
appears in the numerator, the residue is zero). Reading out the residue
we get
\eqn\spirurfour{\eqalign{
C_{34}^{(1)}  = & -{ [4~5] \gb{5|P_{345}|6} \gb{3|P_{345}|6}^2
\over \vev{4~5} [6~1][1~2] \gb{5|P_{345}|5} \gb{5|P_{345}|2} P_{345}^2}
}}
Note here that the result of the integration is the residue with an extra
minus sign. However, the coefficient of a bubble  will be just
 the sum of the residues at the poles.

\subsubsec{The Results}

Since for the other integrations, the procedure is exactly same as
above, we list our results directly. The coefficient in the cut
$C_{612}$ is given by
\eqn\adjcentone{\eqalign{
c_{2:3;6}
 = & -{ \gb{3|P_{612}|6}^2 \over [6~1][1~2] \vev{3~4}
\vev{4~5} \gb{5|P_{612}|2} P_{612}^2} \left(
{ \gb{3|P_5|6} P_{612}^2 \over \gb{5|P_{612}|5}}
+{\gb{3|P_{612} P_2 P_{612} |6} \over \gb{2|P_{612}|2}}\right)
}}
The  coefficient in the cut
$C_{234}$ is given by
\eqn\adjcenttwo{\eqalign{
c_{2:3;2}
= & -{  \gb{1|P_{561}|4}^2 \over \vev{5~6}\vev{6~1}
[2~3][3~4] \gb{5|P_{561}|2} P_{561}^2} \left(
{ \gb{1|P_2|4}P_{561}^2 \over \gb{2|P_{561}|2}}+
{\gb{1|P_{561}P_5 P_{561}|4} \over \gb{5|P_{561}|5}} \right)
}}
The   coefficient in the cut
$C_{34}$ is given by
\eqn\adjcentthree{\eqalign{
c_{2:2;3}
 = & { [4~5] \gb{5|P_{345}|6} \gb{3|P_{345}|6}^2
\over \vev{4~5} [6~1][1~2] \gb{5|P_{345}|5} \gb{5|P_{345}|2} P_{345}^2}
+{ \vev{2~3} \gb{1|P_{234}|2} \gb{1|P_{234}|4}^2
\over [2~3] \vev{5~6}\vev{6~1} \gb{2|P_{234}|2} \gb{5|P_{234}|2}
P_{234}^2}
}}
Finally, the  coefficient in the cut
$C_{61}$ is given by
\eqn\adjcentfour{\eqalign{
c_{2:2;6}
= & { [5~6] \gb{5|P_{561}|4} \gb{1|P_{561}|4}^2 \over
\vev{5~6} [2~3] [3~4] \gb{5|P_{561}|2} \gb{5|P_{561}|5} P_{561}^2}
+{ \vev{1~2} \gb{3|P_{612}|2} \gb{3|P_{612}|6}^2 \over
[1~2] \vev{3~4}\vev{4~5} \gb{5|P_{612}|5} \gb{2|P_{612}|2} P_{612}^2}
}}
It is easy to check that the sum of all these coefficients is
equal to the tree-level amplitude as required by the divergent
behavior discussed in Section 2.

\subsubsec{Comparison with Known Results:}

The same amplitude has been calculated in \BidderTX, where the result was given
by
\eqn\DixonA{\eqalign{
A  = & a_1 {\rm K}_0[s_{61}]+ a_2 {\rm K}_0[s_{34}] -{1\over 2}
\left[ b_1{{\rm L}_0[s_{345}/s_{61}] \over s_{61}}+b_2
{{\rm L}_0[s_{234}/s_{34}]\over s_{34}} \right. \cr &  \left.
+ b_3 {{\rm L}_0[s_{234}/s_{61}]\over s_{61}}
+b_4 {{\rm L}_0[s_{345}/s_{34}]\over s_{34}} \right]
}}
with
\eqn\DixonAcoeff{\eqalign{
a_1  = & a_2 = {1\over 2} A^{\rm tree} \cr
b_1  = & { \gb{ 3|P_{345}|6}^2 \gb{3|P_{345}P_{345}2-P_{345}2 P_{345}|6}
\over \gb{5|P_{345}|2} [6~1] [1~2]\braket{3~4}\braket{4~5} P_{345}^2}\cr
b_4  = & { \gb{ 3|P_{345}|6}^2 \gb{3|-5 P_{345}P_{345}+P_{345}5
 P_{345}|6}
\over \gb{5|P_{345}|2} [6~1] [1~2]\braket{3~4}\braket{4~5} P_{345}^2}\cr
b_2  = & { \gb{1|P_{234}|4}^2 \gb{1|P_{234}2P_{234}-P_{234}P_{234}2|4}
\over \gb{5|P_{234}|2} [2~3][3~4] \braket{5~6}\braket{6~1} P_{234}^2}\cr
b_3  = & { \gb{1|P_{234}|4}^2 \gb{1|5P_{234}P_{234}-P_{234}5P_{234}|4}
\over \gb{5|P_{234}|2} [2~3][3~4] \braket{5~6}\braket{6~1} P_{234}^2}
}}

The functions ${\rm K}_0$ and ${\rm L}_0$ are defined by \BernCG\
\eqn\quaint{
{\rm K}_0(s)={1 \over \epsilon (1-2\epsilon)}(-s)^{-\epsilon}, \qquad
{\rm L}_0(r)={\ln(r) \over 1-r}.
}
The function ${\rm K}_0$ is proportional to the bubble integral, and the function ${\rm L}_0$ is related to the Feynman parameter integral for a two-mass triangle integral.
These two functions are in fact related by the identity
\eqn\sotoo{{{\rm L}_0[s_{1}/s_{2}] \over s_{2}}  =  {{\rm K}_0[s_{2}]-{\rm K}_0[s_{1}]
\over  s_{2}-s_{1}} + \O(\epsilon).
}
Therefore \DixonA\ can be brought to the following form:
\eqn\DixonB{\eqalign{
A = & \left(a_1 -{1\over 2} {b_1\over  s_{61}-s_{345}}
-{1\over 2} {b_3\over s_{61}-s_{234}} \right) {\rm K}_0[s_{61}]+
\left( a_2-{1\over 2} {b_2\over  s_{34}-s_{234}}-{1\over 2}
{b_4\over s_{34}-s_{345}}\right) {\rm K}_0[s_{34}] \cr
&  +{1\over 2}\left[ \left( {b_1\over  s_{61}-s_{345}}
+{b_4 \over s_{34}-s_{345}}\right) {\rm K}_0[s_{345}] +
\left( {b_2\over s_{34}-s_{234}}+{b_3 \over s_{61}-s_{234}}\right)
{\rm K}_0[s_{234}] \right]
}}
Each quantity in parentheses corresponds to one of the bubble coefficients in \adjcentone-\adjcentfour.
It is easy to check that our results agree with \DixonB.

\subsec{Second Configuration: $A(1^-,2^-,3^+,4^-,5^+,6^+)$}

This configuration has the following $\QZ_2$ symmetry:
$P_\a: i\leftrightarrow 7-i$ plus conjugation. With this
helicity assignment, we have box, triangle and bubble contributions.
The box part is easy to calculate by quadruple cuts.

The nonzero box contributions come from both  two-mass hard
and one-mass box integrals. For the two-mass hard box integrals
$I^{2m~h}_{4:2;i}$, $i=2,4,6$, the coefficients are
\eqn\Secboxcoeff{\eqalign{
c^{2m~h}_{4:2;2}  = & -{P_{61}^2 P_{456}^2\over 2}
{ \gb{4|P_{456}|3}^2  \braket{4~6}[3~1]
\over \gb{6|P_{456}|1}^2 \braket{4~5}\braket{5~6}[1~2][2~3]}\cr
c^{2m~h}_{4:2;4}  = & -{P_{23}^2 P_{612}^2 \over 2}
{ [2~6] \gb{4|P_{612}|2} \gb{4|P_{612}|6}^2
\over \braket{4~5} [6~1][1~2] \gb{5|P_{612}|2}\gb{3|P_{612}|2}^2} \cr
c^{2m~h}_{4:2;6}
  = & -{P_{45}^2 P_{561}^2 \over 2}
{\braket{1~5}\gb{1|P_{561}|3}^2 \gb{5|P_{561}|3}
 \over [2~3]\braket{5~6}\braket{6~1}
\gb{5|P_{561}|2} \gb{5|P_{561}|4}^2}
}}
It is easy to see that $c^{2m~h}_{4:2;4}$ and $ c^{2m~h}_{4:2;6}$
 are mapped to each other under
$P_\a$ while $c^{2m~h}_{4:2;2}$ is invariant. For the one-mass
box integrals $I^{1m}_{4;i}$, $i=5,6$, the coefficients are
\eqn\Secboxcoefftwo{\eqalign{
c^{1m}_{4;5}  = & -{P_{23}^2 P_{34}^2\over 2}{  \gb{1|P_{234}|2} \gb{1|P_{234}|3}^2
\over \braket{5~6}\braket{6~1} [4~2]^2 \gb{5|P_{234}|2} P_{234}^2} \cr
c^{1m}_{4;6}  = & -{P_{34}^2 P_{45}^2 \over 2}{ \gb{5|P_{345}|6} \gb{4|P_{345}|6}^2
\over [6~1][1~2] \braket{3~5}^2 \gb{5|P_{345}|2} P_{345}^2}
}}
which are mapped to each other under $P_\a$.

For the triangle part, as we argued in general, we need only pay
attention to the three-mass triangle part, $I^{3m}_{3}$. For this
case, there is only one $I^{3m}_{3}$ function with the distribution
$(23|45|61)$. We can calculate the coefficient by triple cut
in principle, but we choose to read it out by a corresponding double
cut integration which we will evaluate presently.

For the bubble part, we have following cuts: three particle channels
$C_{123}$, $C_{612}$ and $C_{234}$; two particle channels
$C_{23}$, $C_{34}$, $C_{45}$ and $C_{61}$. Among them,
the pairs
$(C_{612},C_{234})$ and $( C_{23}, C_{45})$
are exchanged under $\QZ_2$ symmetry
while others are invariant. So in total we have five independent
double cuts with the following integrands.
\eqn\SecbubbleInt{\eqalign{
C_{123}  = & -\int d\mu { \gb{4|P_{456}|3}^2  \over [1~2][2~3]
\braket{4~5}\braket{5~6} P_{456}^2} { [3~\ell_1] \braket{\ell_1~4}
\over [1~\ell_1] \braket{\ell_1~6}} \cr
C_{234}  = &  -\int d\mu { \gb{1|P_{561}|3}^2 \over [2~3][3~4]
\braket{5~6}\braket{6~1} P_{561}^2}{[3~\ell_2]\braket{\ell_2~1}[3~\ell_1]
\over  [4~\ell_2]\braket{\ell_2~5}[2~\ell_1]} \cr
C_{23}  = & \int d\mu \left(
{ [5~6]^4 \braket{4~2}^2 \over \braket{2~3}[5~6][6~1]
P_{561}^2 \gb{4|P_{561}|1}} { \braket{4~\ell_1}[\ell_1~3]\over
[2~\ell_1]\gb{\ell_1|P_{561}|5}} \right. \cr
&  - {\gb{4|P_{456}|3}^2 \over [2~3] \braket{4~5}\braket{5~6}
P_{456}^2 \gb{4|P_{456}|1}} {\gb{4|P_{456}|\ell_2} \braket{\ell_2~2}
\braket{\ell_2|P_{23}P_{456}|4} \over
\gb{6|P_{456}|\ell_2} \braket{\ell_2~3}
 \gb{\ell_2|P_{23}|1}}\cr
& \left. + {1\over [4~5]\braket{6~1}\braket{2~3}}
{ \gb{1|\ell_2+4+5|5} \over (\ell_2+4+5)^2} {[3~\ell_1]\over [2~\ell_1]}
{\gb{\ell_1|P_{23}|5} \over \gb{\ell_1|P_{23}|4}}
{ ( \gb{1|6|5}\braket{2~\ell_1}+\gb{2|3|5}\braket{1~\ell_1})^2
\over \gb{\ell_1|P_{61}|5} \braket{6|P_{45}P_{23}|\ell_1}} \right)\cr
C_{34}  = &  \int d\mu \left({ [3~4] \gb{4|P_{612}|6}^2
\over [6~1][1~2] P_{34}^2
P_{612}^2 \gb{5|P_{612}|2} } { [3~\ell_2] \gb{\ell_2|P_{612}|6}
\over [4~\ell_2] \braket{\ell_2~5}} \right. \cr & \left. +
{\braket{3~4} \gb{1|P_{561}|3}^2 \over \braket{5~6}\braket{6~1} P_{34}^2
P_{561}^2 \gb{5|P_{561}|2}} { \braket{\ell_1~4} \gb{1|P_{561}|\ell_1}
\over \braket{\ell_1~3} [2~\ell_1]}\right) \cr
C_{61}= & \int d\mu \left( {[1~6] \braket{1~2}^2 [3~5]^4 \over [3~4][4~5]
\gb{2|P_{345}|5} P_{61}^2 P_{345}^2} {[6~\ell_2] \braket{2~\ell_1}
\over [1~\ell_2] \gb{\ell_1|P_{345}|3}} \right. \cr &
+{ \braket{6~1} [5~6]^2 \braket{4~2}^4 \over \braket{2~3}\braket{3~4}
\gb{2|P_{234}|5} P_{61}^2 P_{234}^2}{\braket{\ell_1~1} [5~\ell_2]
\over \braket{\ell_1~6}\gb{4|P_{234}|\ell_2}}  \cr
& + \left. {1\over \braket{6~1} [2~3] \braket{4~5}}
{ \gb{4|\ell_2+2+3|3} \over (\ell_2+2+3)^2}{  \gb{\ell_1|P_{61}|3}
\braket{4~\ell_1}\braket{1~\ell_1} ( \gb{4|P_{12}|3} \braket{1~\ell_1}
-\braket{4~1}\gb{\ell_1|6|3})^2
\over \braket{\ell_1|P_{61}P_{23}|4} \braket{\ell_1~5}\braket{\ell_1~6}
\gb{\ell_1|P_{61}|2} \gb{\ell_1|P_{45}|3}} \right)
}}
Again, when we calculate integrands we need to use the tree-level
amplitude of four gluons and a pair of fermions or scalars given
in Appendix B.

Now we will perform the integration. Compared to the previous
subsection, this integration is more involved.   We  use two
examples to demonstrate our method.

\subsubsec{The Cut $C_{234}$}

The integral in \SecbubbleInt\ may be reduced to
\eqn\Cttfint{\eqalign{
C_{234} = & C \int \braket{\ell~d\ell}[\ell~d\ell] { P_{234}^2
\over \gb{\ell|P_{234}|\ell}^2} { [3~\ell] \over [4~\ell]}
{\braket{\ell~1}  \gb{\ell|P_{234}|3} \over \braket{\ell~5}
\gb{\ell|P_{234}|2}}
}}
after the $t$ integration, where $\ell=\ell_2$ and
$$C=-{ \gb{1|P_{234}|3}^2 \over [2~3][3~4]
\braket{5~6}\braket{6~1} P_{234}^2}.$$
Now we can see the new feature in the above
integrand: it depends on  the antiholomorphic variable $|\ell]$ as well as
the holomorphic variable $\ket{\ell}$. To simplify the calculation,
we split the above integrand by multiplying numerator and denominator by
$\gb{\ell|P_{234}|4}$. Then use a Schouten identity to
rewrite $[3~\ell]\gb{\ell|P_{234}|4}=[3~4]\gb{\ell|P_{234}|\ell}
-[3|P_{234}|\ell\rangle [\ell~4]$, we get
\eqn\Cttfinttwo{\eqalign{
C_{234} = & C P_{234}^2\int \braket{\ell~d\ell}[\ell~d\ell]
{\braket{\ell~1}  \gb{\ell|P_{234}|3} \over \braket{\ell~5}
\gb{\ell|P_{234}|2}\gb{\ell|P_{234}|4}}
\left( {[3~4]\over [4~\ell]\gb{\ell|P_{234}|\ell}}
+{ \gb{\ell|P_{234}|3} \over \gb{\ell|P_{234}|\ell}^2}\right) \cr
\equiv & C_{234}^{(1)} + C_{234}^{(2)}
}}

Splitting the whole integral into two pieces not only
simplifies the calculation  but also provides a nice way to
separate the various contributions. As we will see shortly,
the first term will produce a pure logarithmic contribution which
is related to the imaginary part of the box integral while the
second term produces a rational function which is exactly the
coefficient of the bubble function.
This same pattern will show up in every calculation we meet. Since
we have already found the coefficient of the box integral from a quadruple cut,  we can neglect this term and concentrate on the rational piece only.
In other words, this splitting is the canonical way to separate
box, triangle and bubble contributions.

Now let us do the integration term by term. For $C_{234}^{(2)}$
we have
\eqn\Cttfintrat{\eqalign{
C_{234}^{(2)} = &  C P_{234}^2\int \braket{\ell~d\ell}[\ell~d\ell]
{\braket{\ell~1}  \gb{\ell|P_{234}|3} \over \braket{\ell~5}
\gb{\ell|P_{234}|2}\gb{\ell|P_{234}|4}}
{ \gb{\ell|P_{234}|3} \over \gb{\ell|P_{234}|\ell}^2} \cr
= &  - C P_{234}^2 \left( {\braket{\ell~1}  \gb{\ell|P_{234}|3}^2
[\eta~\ell] \over \braket{\ell~5}
\gb{\ell|P_{234}|2}\gb{\ell|P_{234}|4} \gb{\ell|P_{234}|\eta}
\gb{\ell|P_{234}|\ell}}\right)_{\rm pole}
}}
where $(~~)_{\rm pole}$ means to sum the residues of all poles,
which is the consequence of the integration
$\int \braket{\ell~d\ell}[\ell~d\ell]$. Choosing
$|\eta]=|P_{234}|1\rangle$, we find that there are three poles
giving nontrivial contributions: $\ket{\ell}=\ket{5}$,
$\ket{\ell}=|P_{234}|2]$ and  $\ket{\ell}=|P_{234}|4]$. Summing up
these contributions we finally get
\eqn\Cttfintrattwo{\eqalign{
C_{234}^{(2)} = &  { \gb{1|P_{234}|3}^2 \over [2~3][3~4]\braket{5~6}
\braket{6~1} P_{234}^2} \left(-{ \gb{1|P_6|5} \gb{5|P_{234}|3}^2 \over
\gb{5|P_{234}|2}
\gb{5|P_{234}|4} \gb{5|P_{234}|5}} \right.  \cr
&  \left. + { \braket{1~2} [2~3]^2 P_{234}^2
\over  [2~4]\gb{5|P_{234}|2} \gb{2|P_{234}|2}}
 + { \braket{1~4} [3~4]^2 P_{234}^2\over [4~2]\gb{5|P_{234}|4}
\gb{4|P_{234}|4}} \right)
}}

For $C_{234}^{(1)}$
we have
\eqn\Cttfintlog{\eqalign{
C_{234}^{(1)} = &  C P_{234}^2\int \braket{\ell~d\ell}[\ell~d\ell]
{\braket{\ell~1}  \gb{\ell|P_{234}|3} \over \braket{\ell~5}
\gb{\ell|P_{234}|2}\gb{\ell|P_{234}|4}}
{[3~4]\braket{\ell~4}\over \gb{\ell|p_4|\ell}\gb{\ell|P_{234}|\ell}}
\cr
 = &  C P_{234}^2\int_0^1 dz\int \braket{\ell~d\ell}[\ell~d\ell]
{\braket{\ell~1}  \gb{\ell|P_{234}|3} [3~4]\braket{\ell~4}
\over \braket{\ell~5}
\gb{\ell|P_{234}|2}\gb{\ell|P_{234}|4}}{1 \over
\gb{\ell|P|\ell}^2},
}}
where in the second line we have used the Feynman parametrization
to rewrite the integrand with $P=z P_{4561}-p_4$. At this stage,
the integration is easy to do and given by
\eqn\Cttfintlogtwo{\eqalign{
C_{234}^{(1)} = & - C P_{234}^2 \int_0^1 dz \left(
{  \gb{\ell|P_{234}|3} [3~4]\braket{\ell~1}
\over \braket{\ell~5}
\gb{\ell|P_{234}|2}\gb{\ell|P_{234}|4}} { \gb{4|P|\ell}
\over P^2  \gb{\ell|P|\ell}} \right)|_{\rm pole} \cr
= & C
P_{234}^2 \int_0^1 dz \left({ \gb{5|P_{234}|3}[3~4] \braket{1~5}
\over \gb{5|P_{234}|2}\gb{5|P_{234}|4}}{ \gb{4|P|5}\over P^2
\gb{5|P|5} } + { [2~3][3~4]\gb{1|P_{234}|2} \over \gb{5|P_{234}|2}[2~4]}
{ \braket{4|P P_{234}|2} \over P^2 \gb{2|P_{234} P P_{234}|2} }
\right. \cr & \left.
-{ [3~4]^2 \gb{1|P_{234}|4} \over [4~2] \gb{5|P_{234}|4}}
{ \braket{4|P P_{234}|4} \over P^2 \gb{4|P_{234} P P_{234}|4} }
\right) \cr
 = & - { \braket{1~5}\gb{1|P_{234}|3}^2 \gb{5|P_{234}|3}
 \over [2~3]\braket{5~6}\braket{6~1}
\gb{5|P_{234}|2} \gb{5|P_{234}|4}^2} \left(
\ln { P_{234}^2 \over P_{234}^2-P_{23}^2} +\ln { -P_{45}^2 \over
P_{234}^2- P_{61}^2}\right)\cr
& +  {  \gb{1|P_{234}|2} \gb{1|P_{234}|3}^2
\over \braket{5~6}\braket{6~1} [4~2]^2 \gb{5|P_{234}|2} P_{234}^2}
 \ln \left( { P_{23}^2 P_{34}^2 \over (P_{234}^2-P_{23}^2)
(P_{234}^2 -P_{34}^2)} \right)
}}
It is easy to see that the first logarithm is the contribution
of two-mass-hard box integral  $I^{2m~h}_{4:2;6}$ and the second logarithm
is the contribution of the one-mass box function  $I^{1m}_{4;5}$.

\subsubsec{The Cut $C_{23}$}

Now we consider the cut $C_{23}$ from which we can read out the
coefficient of triangle function. The integrand is given by
three terms. For the first two terms, the calculations are similar to those above,
so we just list the results for coefficients of bubbles
(and neglect those of boxes). They are
\eqn\Cutdifone{\eqalign{
c_{2:2;2} =& c_{2:2;2}^{(1)} + c_{2:2;2}^{(2)} + c_{2:2;2}^{(3)}; \cr
c_{2:2;2}^{(1)}  = & {\braket{2~4}^2 [5~6]^4 \over [5~6][6~1]
P_{561}^2} {\gb{4|P_{561}|5} \over \gb{4|P_{561}|1}}
{\gb{5|P_{561}|3} \over \gb{3|P_{561}|5}} {1\over
\gb{5|P_{561} P_{23} P_{561}|5}} \cr
c_{2:2;2}^{(2)}  = &
{\gb{4|P_{456}|3}^2 \over [2~3] \braket{4~5}\braket{5~6}
P_{456}^2 \gb{4|P_{456}|1}} \left( - {[3~2] \braket{2~1}
\gb{4|P_{456}|1}^2 \over [2~1]\gb{6|P_{456}|1} \gb{1|P_{23}|1}}
\right. \cr
 & \left. + {  \braket{4~6}^2 [2~3] \gb{2|P_{456}|6}
(P_{456}^2)^2 \over \gb{6|P_{456}|2} \gb{6|P_{456}|1}
\gb{6|P_{456} P_{23} P_{456}|6}} \right)
}}

The third term is a little involved.  Defining
\eqn\Cutdifdef{\eqalign{
g(\ell)  = &
-{\gb{\ell|P_{23}|5} \over \gb{\ell|P_{23}|4}}
{ ( \gb{1|6|5}\braket{2~\ell}+\gb{2|3|5}\braket{1~\ell})^2
\over \gb{\ell|P_{61}|5} \braket{\ell|P_{23} P_{45}|6}}\cr
Q  = & {P_{61}^2 P_{23}+   P_{23}^2 P_{61} \over  P_{23}^2},~~~~~
Q^2={ P_{61}^2 P_{45}^2 \over  P_{23}^2}
}}
it is easy to see that the third integration becomes
(with $C= {1\over [4~5]\braket{6~1}\braket{2~3}}$)
\eqn\Cutdifthirdone{\eqalign{
C_{23}^{(3)} = &  -C \int \braket{\ell~d\ell}[\ell~d\ell]
{ P_{23}^2 \over \gb{\ell|P_{23}|\ell}^2}
{[3~\ell]\over [2~\ell]}
{ \gb{1|P_{61}|5} \gb{\ell|P_{23}|\ell} + P_{23}^2
\gb{1|\ell|5} \over \gb{\ell|Q|\ell} P_{23}^2}  g(\ell)
}}
after the $t$-integration. Now we can use our method to
split the antiholomorphic part in the denominator as
\eqn\Cutdifthirdtwo{\eqalign{
C_{23}^{(3)} = &  C \int \braket{\ell~d\ell}[\ell~d\ell]
\left[
{1\over \gb{\ell|P_{23}|\ell}[2~\ell]}{ g(\ell) [3~2]\over
\gb{\ell|Q|2}}\left( -\gb{1|6|5} +
{\braket{3~2}\braket{1~\ell}[2~5]\over \braket{\ell~3}}\right)\right.\cr
&  - {1\over \gb{\ell|P_{23}|\ell}^2} {  g(\ell) P_{23}^2
\braket{1~\ell} \gb{\ell|P_{23}|5} \braket{\ell~2}\over
\braket{\ell~3}\braket{\ell|P_{23} Q|\ell}}\cr
 &\left. - {1\over
\gb{\ell|P_{23}|\ell} \gb{\ell|Q|\ell}}{  g(\ell) \gb{\ell|Q|3}
\over \gb{\ell|Q|2} \braket{\ell|P_{23} Q|\ell}} \left(
\gb{1|6|5}\braket{\ell|P_{23} Q|\ell}+ P_{23}^2
\braket{1~\ell}\gb{\ell|Q|5} \right) \right]
}}
The first, second and third lines of \Cutdifthirdtwo\
are respectively the contributions from the box, bubble, and triangle.
For the second term,
it is easy to read out
\eqn\Cutdifthirdthree{\eqalign{
c_{2:2;2}^{(3)} = & -{1\over [4~5]\braket{6~1}}
\left( {  g(\ell)
\braket{1~\ell} \gb{\ell|P_{23}|5} [3~\ell]\over
\braket{\ell~3}\braket{\ell|P_{23} Q|\ell} \gb{\ell|P_{23}|\ell}}\right)_{\rm pole}
}}
There are five poles giving non-zero contributions:
three from $g(\ell)$ and two from the factor
$\braket{\ell|P_{23} Q|\ell}$. To find the location of the last
two poles, we use the following method. Taking two arbitrary
external momenta $\ket{a}$ and $\ket{b}$, since the
spinor is two dimensional, we can represent  $\ket{\ell}$ as
\eqn\unkown{ \ket{\ell}=(\ket{a}+ x\ket{b})}
with undetermined
complex variable $x$.\foot{In principle we should write $\ket{\ell}=\a (\ket{a}+ x\ket{b})$, but one can check that the factor $\a$ drops out of the final expression, so we can set $\a=1$.}
Putting it
back into $\braket{\ell|P_{23} Q|\ell}$, we get a quadratic
equation whose solutions are
\eqn\unkownx{ x_\pm={ -(\braket{a|P_{23} Q|b}+\braket{b|P_{23} Q|a})\pm
 \braket{a~b} \sqrt{\Delta_{3m}}\over 2
\braket{b|P_{23}Q|b}}}
where
\eqn\unkownPQ{\Delta_{3m}  =   (P_{61}^2)^2+(P_{45}^2
)^2+(P_{23}^2)^2-2 P_{61}^2 P_{23}^2 -2 P_{23}^2 P_{45}^2 -2
P_{45}^2 P_{61}^2}
Now we can write the residue as
\eqn\Cutdifthirdfour{\eqalign{
c_{2:2;2}^{(3)} = & -{1\over [4~5]\braket{6~1}}\sum_{i=1}^5
\left( \braket{\ell~\ell_i}{  g(\ell)
\braket{1~\ell} \gb{\ell|P_{23}|5} [3~\ell]\over
\braket{\ell~3}\braket{\ell|P_{23} Q|\ell}\gb{\ell|P_{23}|\ell}
}\right)_{\ell\to \ell_i}
}}
with the following five poles:
\eqn\fivepoles{\eqalign{
\ket{\ell_1}= & |P_{23}|4],~~~\ket{\ell_2}=|P_{61}|5],~~~~\ket{\ell_3}=
|P_{23} P_{45}|6\rangle \cr
\ket{\ell_4}= & \ket{a}+ x_+\ket{b},~~~~~
\ket{\ell_5}=  \ket{a}+ x_-\ket{b},
}}
It does not seem useful to write the
expression \Cutdifthirdfour\
explicitly because it is rather long and complicated.
However, its structure is very clear and easy to implement in a computer program.
One has to evaluate
the right hand side of \Cutdifthirdfour\ for the five values of
$|\ell\rangle$
given by \fivepoles\ and add up the obtained contributions.
We find that the \Cutdifthirdfour\ is the most convenient
way to present the answer.

It is worth remarking that although the square root shows up in
above expression, the final result $c_{2:2;2}^{(3)}$ is rational.
A similar feature is encountered in calculating
the coefficient of a four-mass box integral from a quadruple cut \BrittoNC.

Now we are left with the third term of  $C_{23}^{(3)}$ which can
be expressed as
\eqn\thirdthirdone{\eqalign{
C_{23}^{(3;3)} = & - C \int_0^1 dz \int \braket{\ell~d\ell}[\ell~d\ell]
{1\over
\gb{\ell|P|\ell}^2 }{  g(\ell) \gb{\ell|Q|3}
\over \gb{\ell|Q|2} \braket{\ell|P_{23} Q|\ell}} \left(
\gb{1|6|5}\braket{\ell|P_{23} Q|\ell}+ P_{23}^2
\braket{1~\ell}\gb{\ell|Q|5} \right)
}}
with $P\equiv (1-z)P_{23}+z Q$. Do the integration
$\int \braket{\ell~d\ell}[\ell~d\ell]$, and  we are left with
\eqn\thirdthirdtwo{\eqalign{
C_{23}^{(3;3)} = &  C \int_0^1 dz
\left[{[\eta~\ell]\over \gb{\ell|P|\ell} \gb{\ell|P |\eta}}
{  g(\ell) \gb{\ell|Q|3}
\over \gb{\ell|Q|2} \braket{\ell|P_{23} Q|\ell}} \left(
\gb{1|6|5}\braket{\ell|P_{23} Q|\ell}+ P_{23}^2
\braket{1~\ell}\gb{\ell|Q|5} \right)\right]_{\rm pole}\cr
= &  C \int_0^1 dz
\left[{\gb{\W \eta|P |\ell}\over \gb{\ell|P|\ell} P^2}
{  g(\ell) \gb{\ell|Q|3}
\over \braket{\ell~\W \eta}
\gb{\ell|Q|2} \braket{\ell|P_{23} Q|\ell}} \left(
\gb{1|6|5}\braket{\ell|P_{23} Q|\ell}+ P_{23}^2
\braket{1~\ell}\gb{\ell|Q|5} \right)\right]_{\rm pole}
}}
where at the second line we have chosen $|\eta]=|P\ket{\W \eta}$ with
arbitrary $\W \eta$. The advantage of this choice is that
now every pole is independent of the Feynman parameter
$z$, and we can evaluate $\int_0^1 dz$ before taking residues of
poles. The integration is of the following pattern,
\eqn\intpattern{\eqalign{
I_f  \equiv & \int_0^1 dz { ( z c_1 +c_2) \over ( a_0 z^2 + a_1 z+ a_2)
( z b_1+ b_2)} \cr
 = & \int_0^1 dz   { b_1( -b_2 c_1+ b_1 c_2)
\over ( a_2 b_1^2 - a_1 b_1 b_2+ a_0 b_2^2)} {1\over
( z b_1+ b_2)}  \cr
& + \int_0^1 dz
 { ( b_2 c_1-b_1 c_2)\over  2( a_2 b_1^2 - a_1 b_1 b_2+ a_0 b_2^2)}
{ (2 z a_0 + a_1) \over ( a_0 z^2 + a_1 z+ a_2)} \cr
& + \int_0^1 dz{   (2 a_2 b_1 c_1-a_1 b_2 c_1-a_1 b_1 c_2+
2 a_0 b_2 c_2) \over 2( a_2 b_1^2 - a_1 b_1 b_2+ a_0 b_2^2)}
{1\over ( a_0 z^2 + a_1 z+ a_2)}
}}
where
\eqn\intpara{\eqalign{
a_0  = & (Q-P_{23})^2,~~~~~a_1= 2 P_{23} \cdot (Q-P_{23}),~~~~~
a_2= P_{23}^2 \cr
b_1  = & \gb{\ell| (Q-P_{23})|\ell},~~~~b_2=\gb{\ell|P_{23}|\ell}
~~~~c_1=\gb{\W \eta|(Q-P_{23})|\ell},~~~~c_2=\gb{\W \eta|P_{23}|\ell}
}}
We have split $I_f$ into three terms. Among them, the first two
terms give the imaginary part of the box integral while the last
term is exactly the cut contribution of the three-mass triangle function.
Let us define the function
\eqn\Rfun{R_1 (a_j,b_j,c_j) =
{   (2 a_2 b_1 c_1-a_1 b_2 c_1-a_1 b_1 c_2+
2 a_0 b_2 c_2) \over 2( a_2 b_1^2 - a_1 b_1 b_2+ a_0 b_2^2)}.
}
We would like to point out that there is an important subtlety with the above 
procedure.\foot{This subtlety was overlooked in previous versions of this
paper. We would like to thank P. Mastrolia for stimulating discussions pointing to this issue and R. K. Ellis who independently found the problem and
informed us.}
It is correct only if the denominator $a_2 b_1^2 - a_1 b_1 b_2+ a_0 b_2^2$
does not vanish. 
However, it can be shown that this denominator vanishes for $\ket{\ell}$
satisfying $\braket{\ell|P_{23} Q|\ell}=0$ which is one of the poles in 
eq.~\thirdthirdone.
This means that we have to redo the simple pole expansion assuming that 
$ a_2 b_1^2 - a_1 b_1 b_2+ a_0 b_2^2 =0$.
In fact, we need only the term proportional to ${1 \over (a_0z^2+a_1z+a_2)}$
because only such a term contributes to the three-mass triangle upon the $z$-integration. 
Redoing the simple fraction expansion we find that ${1 \over (a_0z^2+a_1z+a_2)}$
is multiplied by the function $R_2(a_j,b_j,c_j)$, where 
\eqn\Rtwo{
R_2(a_j,b_j,c_j) = {(a_2b_2c_1 +a_2b_1c_2 -a_1b_2c_2) \over b_2(2a_2b_1-a_1b_2)}}
and all $a$'s, $b$'s and $c$'s are as before. 
Thus,
we finally have the coefficient of three-mass triangle as
\eqn\threemassCoe{\eqalign{
& c^{3m}_{3:2:2;2}  = {1\over [4~5]\braket{6~1}\braket{2~3}} \times
\cr
& ~~~ \sum_{i=1}^5\left[ \braket{\ell~\ell_i}{  g(\ell) \gb{\ell|Q|3}
R_1(a_j,b_j,c_j)\over \braket{\ell~\W \eta}
\gb{\ell|Q|2} \braket{\ell|P_{23} Q|\ell}} \left(
\gb{1|6|5}\braket{\ell|P_{23} Q|\ell}+ P_{23}^2
\braket{1~\ell}\gb{\ell|Q|5} \right)\right]_{\ell\to \ell_i}
\cr
& ~~~ + {1\over [4~5]\braket{6~1}\braket{2~3}} \times 
\cr 
& ~~~ \sum_{i=6}^7\left[ \braket{\ell~\ell_i}{  g(\ell) \gb{\ell|Q|3}
R_2(a_j,b_j,c_j)\over \braket{\ell~\W \eta}
\gb{\ell|Q|2} \braket{\ell|P_{23} Q|\ell}} \left(
\gb{1|6|5}\braket{\ell|P_{23} Q|\ell}+ P_{23}^2
\braket{1~\ell}\gb{\ell|Q|5} \right)\right]_{\ell\to \ell_i}.
}}
Here, in the first term, the summation is over the poles
\eqn\fivepolesone{\eqalign{
\ket{\ell_1}= & |P_{23}|4],~~~\ket{\ell_2}=|P_{61}|5],~~~~\ket{\ell_3}=
|P_{23} P_{45}|6\rangle, \cr
\ket{\ell_4}= & |Q|2],~~~~\ket{\ell_5}=\ket{\W \eta}.}}
In the second term, the summation is over the remaining two poles
\eqn\fivepolestwo{
\ket{\ell_6}=  \ket{a}+ x_+\ket{b},~~~~~
\ket{\ell_7}=  \ket{a}+ x_-\ket{b}.}
We can choose $\ket{\W \eta}$ properly to reduce the number of poles
further, but we do not do so here. It was checked numerically
that eq.~\threemassCoe\ is indeed independent of $\ket{\W \eta}$.

\subsubsec{The Results}

Now we list  the coefficients of the amplitude  $A(1^-,2^-,3^+,4^-,5^+,6^+)$.
The box coefficients are given by \Secboxcoeff\ and \Secboxcoefftwo.
The three-mass triangle coefficient is given by eq.~\threemassCoe.
For bubbles, we have six coefficients. From cuts in three-particle
channels
we have
\eqn\SecResbubbleone{\eqalign{
c_{2:3;2}
  = & -{ \gb{1|P_{234}|3}^2 \over [2~3][3~4]\braket{5~6}
\braket{6~1} P_{234}^2} \left( { \gb{1|6|5} \gb{5|P_{561}|3}^2 \over
\gb{5|P_{561}|2}
\gb{5|P_{561}|4} \gb{5|P_{561}|5}} \right. \cr
&  \left. + { \braket{1~2} [2~3]^2 P_{561}^2
\over  [2~4]\gb{5|P_{561}|2} \gb{2|P_{561}|2}}
 + { \braket{1~4} [3~4]^2 P_{561}^2\over [4~2]\gb{5|P_{561}|4}
\gb{4|P_{561}|4}} \right) \cr
c_{2:3;6}
  = & -{  \gb{4|P_{612}|6}^2 \over [6~1][1~2] \braket{3~4}
\braket{4~5} P_{612}^2} \left( {\gb{2|1|6} \gb{4|P_{612}|2}^2
\over \gb{5|P_{612}|2} \gb{3|P_{612}|2} \gb{2|P_{612}|2}}
\right. \cr
 & \left. + { [5~6] \braket{4~5}^2 P_{612}^2\over \braket{3~5}
\gb{5|P_{612}|2} \gb{5|P_{612}|5}} +{ [6~3]\braket{3~4}^2
P_{612}^2\over \braket{3~5} \gb{3|P_{612}|2} \gb{3|P_{612}|3}}
\right) \cr
c_{2:3;1}
 = &{ \gb{4|P_{456}|3}^2  \over [1~2][2~3]
\braket{4~5}\braket{5~6} }
\left( { [6~3] \braket{6~4} \over \gb{6|P_{123}|6}
\gb{6|P_{123}|1}} -{ \gb{1|2|3} \gb{4|P_{123}|1}\over  \gb{1|P_{123}|1}
\gb{6|P_{123}|1} P_{123}^2} \right)
}}
From cuts in two-particle channels, we have
\eqn\SecResbubbletwo{
c_{2:2;3}
  =  { [3~5] \gb{5|P_{612}|6} \gb{4|P_{612}|6}^2 \over [6~1][1~2]
\braket{3~5} \gb{5|P_{612}|2} \gb{5|P_{612}|5} P_{612}^2}
+ { \vev{2~4}\gb{1|P_{561}|2} \gb{1|P_{561}|3}^2 \over \braket{5~6}
\braket{6~1}[2~4] \gb{5|P_{561}|2} \gb{2|P_{561}|2} P_{561}^2}
}
and
\eqn\SecResbubblethree{\eqalign{
c_{2:2;2} = & {\braket{2~4}^2 [5~6]^4 \over [5~6][6~1]
P_{561}^2} {\gb{4|P_{561}|5} \over \gb{4|P_{561}|1}}
{\gb{5|P_{561}|3} \over \gb{3|P_{561}|5}} {1\over
\gb{5|P_{561} P_{23} P_{561}|5}}\cr
+ & {\gb{4|P_{456}|3}^2 \over [2~3] \braket{4~5}\braket{5~6}
P_{456}^2 \gb{4|P_{456}|1}} \left( -{[3~2] \braket{2~1}
\gb{4|P_{456}|1}^2 \over [2~1]\gb{6|P_{456}|1} \gb{1|P_{23}|1}}
\right. \cr
 & \left. + {  \braket{4~6}^2 [2~3] \gb{2|P_{456}|6}
(P_{456}^2)^2 \over \gb{6|P_{456}|2} \gb{6|P_{456}|1}
\gb{6|P_{456} P_{23} P_{456}|6}} \right)\cr
- & {1\over [4~5]\braket{6~1}}\sum_{i=1}^5
\left( \braket{\ell~\ell_i}{  g(\ell)
\braket{1~\ell} \gb{\ell|P_{23}|5} [3~\ell]\over
\braket{\ell~3}\braket{\ell|P_{23} Q|\ell}}\right)_{\ell\to \ell_i}
}}
The coefficient $c_{2:2;4}$  can be obtained $c_{2:2;2}$ by a flip symmetry, while
the coefficient $c_{2:2;6}$ can be expressed as $A_{\rm tree}-\sum_{\rm others} c_{2:r;i}$
from the divergence equation \irrel.  We have obtained an analytic expression for $c_{2:2;6}$ using the techniques of this paper and checked numerically that \irrel\ is satisfied.  The analytic expression is rather long, so we omit it here for brevity.

\subsec{Third Configuration: $A(1^-,2^+,3^-,4^+,5^-,6^+)$}

Now we move to the last helicity configuration
$A(1^-,2^+,3^-,4^+,5^-,6^+)$. It has the largest symmetry
$\QZ_6$ generated by $P_\a: i\to i+1$ plus conjugation.
Because of this,  we need to calculate just one coefficient for
each type of function and act on it by $P_\a$ to obtain all the others.
The box coefficients
are
\eqn\Altboxone{
c^{2m~h}_{4:2;1}=  { P_{345}^2 P_{56}^2 \over 2}{ \gb{1|P_{345}|4}^2 \over
\braket{1~2}[3~4] \gb{6|P_{345}|5}^2}
{\gb{1|P_{345}|5}\gb{6|P_{345}|4} \over
\gb{2|P_{345}|5}\gb{6|P_{345}|3}}
}
 and
\eqn\Altboxtwo{
c^{1m}_{4;1}  =  -{P_{45}^2 P_{56}^2\over 2}{ \gb{5|P_{456}|2}^2
\over [1~2][2~3] \braket{4~6}^2 P_{456}^2} { \gb{4|P_{456}|2}
\gb{6|P_{456}|2}\over \gb{4|P_{456}|1} \gb{6|P_{456}|3}}
}

For the triangle integrals, there are two two-mass triangles related by $P_\a$.
For the bubble integrals, the orbit of the cut in a three-particle channel contains three elements,
while the orbit of the cut in a two-particle channel contains six elements.
The representative integration we will perform is the following:
\eqn\Altinte{\eqalign{
C_{123} = & -{ \gb{5|P_{123}|2}^2 \over
[1~2][2~3]\braket{4~5}\braket{5~6} P_{123}^2}
{\gb{5|P_{123}|\ell_2}\gb{\ell_2|P_{123}|2} \over
\gb{6|P_{123}|\ell_2}\gb{\ell_2|P_{123}|1}} {[2~\ell_2]
\braket{5~\ell_2} \over [3~\ell_2] \braket{4~\ell_2}} \cr
C_{12} = & {[4~6]^4 \braket{1~3}^2 [1~2] \over [4~5][5~6]
\gb{3|P_{456}|6} P_{456}^2 P_{12}^2}
{[\ell_1~2] \braket{\ell_1~3} \over [\ell_1~1]\gb{\ell_1|P_{456}|4}}
+{ \braket{3~5}^4 \braket{1~2} [2~6]^2\over \braket{3~4}\braket{4~5}
\gb{3|P_{345}|6}P_{345}^2 P_{12}^2} {\braket{1~\ell_2} [6~\ell_2]
\over \braket{2~\ell_2} \gb{5|P_{345}|\ell_2}}\cr
&  -{1 \over \braket{1~2}[3~4]\braket{5~6}}
{\gb{5|\ell_2+3+4|4}\over (\ell_2+3+4)^2}
{\braket{5~\ell_1}\gb{1|P_{12}|\ell_1} \over \braket{6~\ell_1}
\gb{2|P_{12}|\ell_1}} {\gb{\ell_1|P_{12}|4} \over \braket{5|P_{34}
P_{12}|\ell_1}} { (\gb{5|6|4}\braket{1~\ell_1}+\gb{1|2|4}
\braket{5~\ell_1})^2\over \gb{\ell_1|P_{56}|4} \gb{\ell_1|P_{12}|3}}
}}

Now we perform  the integration for these two cuts.

\subsubsec{The cut $C_{123}$}

After the $t$-integration and splitting we end up with
\eqn\Altthreecutone{\eqalign{
C_{123} = & C{P_{123}^2\gb{5|P_{123}|3}\over \gb{6|P_{123}|3}}
\int \braket{\ell~d\ell}[\ell~d\ell]
{\gb{\ell|P_{123}|2} \braket{\ell~5} \over
\gb{\ell|P_{123}|1} \braket{\ell~4}} {1\over \gb{\ell|P_{123}|3}}
\left( { \gb{\ell|P_{123}|2} \over \gb{\ell|P_{123}|\ell}^2}
+{[2~3]\over \gb{\ell|P_{123}|\ell} [3~\ell]}\right)\cr
&  +  C{P_{123}^2 \braket{5~6}\over \gb{6|P_{123}|3}}
\int \braket{\ell~d\ell}[\ell~d\ell]
{\gb{\ell|P_{123}|2} \braket{\ell~5} \over
\gb{\ell|P_{123}|1} \braket{\ell~4}}{ 1\over \braket{\ell~6}}
\left( - { \gb{\ell|P_{123}|2} \over\gb{\ell|P_{123}|\ell}^2}
+{ \gb{6|P_{123}|2} \over  \gb{\ell|P_{123}|\ell} \gb{6|P_{123}|\ell}}
 \right)
}}
with $C=-{ \gb{5|P_{123}|2}^2 \over
[1~2][2~3]\braket{4~5}\braket{5~6} P_{123}^2}$.
By our standard method, it is clear that among
 these four terms, the first and third terms will give rational
functions  and the second and fourth terms will give  the logarithmic
functions involved in the
box contributions. Carrying out the integration for the first
and third terms we read out the coefficient of bubble $C_{123}$ as
\eqn\Altthreecuttwo{\eqalign{
c_{2:3;1}
 = & -{ \gb{5|P_{123}|2}^2 \over [1~2][2~3]\braket{4~5}\braket{5~6}
P_{123}^2}\left( { \gb{4|P_{123}|2}^2 \braket{4~5}\braket{5~6}[4~6]
\over \gb{4|P_{123}|1}\gb{4|P_{123}|3}\gb{4|P_{123}|4}\braket{4~6}}
\right. \cr
&  \left.  -{ \braket{1~5}[1~2]^2 P_{123}^2 \gb{5|P_{123}|1}
\over [1~3] \gb{4|P_{123}|1}\gb{6|P_{123}|1}\gb{1|P_{123}|1}}
+ {\gb{5|P_{123}|3}\over \gb{6|P_{123}|3}}
{ \braket{5~3} [2~3]^2 P_{123}^2 \over [3~1]
\gb{4|P_{123}|3} \gb{3|P_{123}|3}}\right. \cr
 & \left. +  { \braket{5~6}\over \gb{6|P_{123}|3}}{ \braket{5~4}[4~6]
\gb{6|P_{123}|2}^2 \over \braket{6~4} \gb{6|P_{123}|1}
\gb{6|P_{123}|6}} \right)
}}
We can do similar calculations for the second and fourth terms and
it is easy to check that they
produce the imaginary parts of box integrals.

\subsubsec{The cut $C_{12}$}

Here we consider the cut $C_{12}$. The integrand consists of the three terms.
Integration of the first two terms can be performed by the same procedure as presented earlier
in the paper. Therefore, we will only state the results. The first two terms
contribute to box coefficients, which can be neglected for our purposes, and to the
bubble coefficient $c_{2:2;1}$. The corresponding contributions are
\eqn\eone{
c_{2:2;1}^{(1)}= {[4~6]^4\braket{1~3}^2 \over [4~5][5~6]\gb{3|P_{456}|6}P^2_{456}}
{\gb{3|P_{456}|4} \gb{4|P_{456}|2} \over \gb{2|P_{456}|4} \gb{4|P_{456}P_{12}P_{456}|4}}}
and
\eqn\etwo{
c_{2:2;1}^{(2)}= {\braket{3~5}^4 [2~6]^2 \over \braket{3~4} \braket{4~5}\gb{3|P_{345}|6}P^2_{345}}
{\gb{1|P_{345}|5} \gb{5|P_{345}|6} \over \gb{5|P_{345}|1} \gb{5|P_{345}P_{12}P_{345}|5}}.}
Now we consider the last term. After performing the $t$-integration,
its contribution to the cut $C_{12}$ can be written as follows
\eqn\ethree{
C_{12}^{(3)}=C \int {\braket{\ell~d \ell}[\ell~d\ell] \over \gb{\ell|P_{12}|\ell}^2}
P_{12}^2 g(\ell) {[\ell~2]\over [\ell~1]}
\left(-{\gb{\ell|P_{12}|\ell}\gb{5|6|4}\over P_{12}^2\gb{\ell|Q|\ell}}
+{\braket{5~\ell}[\ell~4]\over \gb{\ell|Q|\ell}} \right),}
where
\eqn\efour{
C= {1\over \braket{1~2} [3~4] \braket{5~6}},}
\eqn\efive{
g(\ell)=-{\braket{\ell~5}\over \braket{\ell~6}}
{\gb{\ell|P_{12}|4} \over \braket{\ell|P_{12}P_{34}|5}}
{(\gb{5|6|4}\braket{1~\ell}+\gb{1|2|4}\braket{5~\ell})^2
\over \gb{\ell|P_{56}|4} \gb{\ell|P_{12}|3}}}
%
%
%
and
\eqn\eseven{
Q={1 \over P_{12}^2}(P_{56}^2P_{12}+P_{12}^2P_{56}).}
Now we use our method to split the integrand in \ethree.
We find that
\eqn\eeight{\eqalign{
C_{12}^{(3)} &= C\int \braket{\ell~d \ell}[\ell~d\ell]
\left[ {1 \over \gb{\ell|P_{12}|\ell}[1~\ell]}
{g(\ell) [1~2] \over \gb{\ell|Q|1}}
\left( \gb{5|6|4} - {\braket{1~2}[4~1]\braket{5~\ell}\over \braket{\ell~2}}
\right) \right. \cr
&- {1 \over \gb{\ell|P_{12}|\ell}^2}
{g(\ell) P_{12}^2 \braket{5~\ell}\gb{\ell|P_{12}|4}\braket{\ell~1}
\over \braket{\ell~2}\braket{\ell|P_{12}Q|\ell}} \cr
&\left. -
{1\over \gb{\ell|P_{12}|\ell} \gb{\ell|Q|\ell}}
{g(\ell)\gb{\ell|Q|2} \over \gb{\ell|Q|1}\braket{\ell|P_{12}Q|\ell}}
(\gb{5|6|4}\gb{\ell|P_{12}Q|\ell}+P_{12}^2\braket{5 \ell}\gb{\ell|Q|4})\right].}}
Among these three parts, the first one contributes to one
of the box coefficients and can be neglected for our purposes.
The second term gives a contribution to the bubble coefficient
$c_{2:2;1}$. Performing the $\ell$-integration, we find this contribution
to be of the following form
\eqn\enine{
c_{2:2;1}^{(3)}=-{1 \over [3~4]\braket{5~6}} \sum_{i=1}^6
\left(\braket{\ell~\ell_i}
{g(\ell) \braket{5~\ell} \gb{\ell|P_{12}|4} [2~\ell] \over
\braket{\ell~2} \braket{\ell|P_{12}Q|\ell} \gb{\ell|P_{12}|\ell}}
\right)_{\ell \to \ell_{i}}.}
Here the locations of the six poles is as follows:
\eqn\eten{\eqalign{
&|\ell_1\rangle =|P_{56}|4], \quad
|\ell_2\rangle =|P_{12}|3], \quad
|\ell_3\rangle =|6 \rangle, \quad
|\ell_4\rangle =|P_{12}P_{34}|5 \rangle, \cr
&|\ell_5\rangle =|a\rangle +x_{-}|b\rangle, \quad
|\ell_6\rangle =|a\rangle +x_{+}|b\rangle,}}
with
\eqn\unkownxtwoagain
{ x_\pm={ -(\braket{a|P_{12} Q|b}+\braket{b|P_{12} Q|a})\pm
 \braket{a~b} \sqrt{\Delta_{3m}}\over 2
\braket{b|P_{12}Q|b}}}
and
\eqn\unkownPQtwoagain{\eqalign{
\Delta_{3m} &   =  \left( (P_{12}^2)^2+(P_{34}^2
)^2+(P_{56}^2)^2-2 P_{12}^2 P_{34}^2 -2 P_{34}^2 P_{56}^2 -2
P_{56}^2 P_{12}^2\right)
}}
Note that the pole $|\ell \rangle =|2\rangle$ does not contribute
because of the factor $[2~\ell]$ in the numerator of \enine.

Now we move on to the last term in \eeight. It will produce
the contribution to the three-mass triangle coefficient. By introducing the
Feynman parameter $z$, we can write it as follows:
\eqn\eeleven{
C_{12}^{(3,3)}  = C \int_{0}^{1}dz \int
{\braket{\ell d~\ell}[\ell~d\ell] \over \gb{\ell|P^2|\ell}}
 {-g(\ell)\gb{\ell|Q|2} \over \gb{\ell|Q|1}\braket{\ell|P_{12}Q|\ell}}
(\gb{5|6|4}\braket{\ell|P_{12}Q|\ell}+P_{12}^2\braket{5~\ell} \gb{\ell|Q|4}),}
where
\eqn\etwelve{
P=(1-z)P_{12}+z Q.}
Performing the $\ell$-integration, we obtain
\eqn\ethirteen{
C_{12}^{(3,3)}=-C\int_{0}^{1} dz \left[
{[\eta~\ell] \over \gb{\ell|P|\ell}\gb{\ell|P|\eta}}
 {-g(\ell)\gb{\ell|Q|2}\over \gb{\ell|Q|1} \braket{\ell|P_{12}Q|\ell}}
(\gb{5|6|4}\braket{\ell|P_{12}Q|\ell}+P_{12}^2\braket{5~\ell} \gb{\ell|Q|4})
\right]_{\rm pole},}
where $|\eta]$ is an arbitrary auxiliary spinor.
Let us write $|\eta]$ as
\eqn\efourteen{
|\eta] =|P|\tilde{\eta}\rangle}
for some spinor $|\tilde{\eta}\rangle$. Then \ethirteen\ becomes
\eqn\efifteen{
C_{12}^{(3,3)}=C\int_{0}^{1} dz \left[
{\gb{\tilde{\eta}|P|\ell} \over \gb{\ell|P|\ell}P^2}
{g(\ell)\gb{\ell|Q|2}\over \braket{\ell~\tilde{\eta}} \gb{\ell|Q|1} \braket{\ell|P_{12}Q|\ell}}
(\gb{5|6|4}\braket{\ell|P_{12}Q|\ell}+P_{12}^2\braket{5~\ell} \gb{\ell|Q|4})
\right]_{\rm pole}.}
The Feynman integral of this type was considered before. It produces
the functions $R_1(a_j, b_j, c_j)$ and $R_2(a_j,b_j,c_j)$. As a result, the contribution
to the three-mass triangle coefficient is given by
\eqn\esixteen{\eqalign{
c_{3:2:2;1}^{3m} =& {1 \over \braket{1~2} [3~4] \braket{5~6}}
\sum_{i=1}^6 \left[\braket{\ell \ell_i} 
{g(\ell)\gb{\ell|Q|2}R_1(a_j, b_j, c_j) 
\over \braket{\ell~\tilde{\eta}} \gb{\ell|Q|1} \braket{\ell|P_{12}Q|\ell}}
(\gb{5|6|4}\braket{\ell|P_{12}Q|\ell}+P_{12}^2\braket{5~\ell} \gb{\ell|Q|4})
\right]_{\ell \to \ell_i} \cr +&
{1 \over \braket{1~2} [3~4] \braket{5~6}}
\sum_{i=7}^8 \left[\braket{\ell \ell_i} 
{g(\ell)\gb{\ell|Q|2}R_2(a_j, b_j, c_j) 
\over \braket{\ell~\tilde{\eta}} \gb{\ell|Q|1} \braket{\ell|P_{12}Q|\ell}}
(\gb{5|6|4}\braket{\ell|P_{12}Q|\ell}+P_{12}^2\braket{5~\ell} \gb{\ell|Q|4})
\right]_{\ell \to \ell_i},
}}
where, in this example, the coefficients $a_j, b_j$ and $c_j$ are
\eqn\eesixteenone{\eqalign{
& a_0=(Q-P_{12})^2, \quad a_1=2P_{12}\cdot (Q-P_{12}), \quad a_2=P_{12}^2, \cr
& b_1 =\gb{\ell|Q-P_{12}|\ell}, \quad b_2 =\gb{\ell|P_{12}|\ell}, \quad
c_1 = \gb{\tilde{\eta}|(Q-P_{12})|\ell}, \quad c_2=\gb{\tilde{\eta}|P_{12}|\ell}.}}
In the first term in eq.~\esixteen, the summation is over the poles 
\eqn\etenprime{\eqalign{
&|\ell_1\rangle =|P_{56}|4], \quad 
|\ell_2\rangle =|P_{12}|3], \quad
|\ell_3\rangle =|6 \rangle, \quad
|\ell_4\rangle =|P_{12}P_{34}|5 \rangle, \cr
&|\ell_5\rangle =|\tilde{\eta} \rangle, \quad
|\ell_6\rangle =|Q|1].}}
In the second term, the summation is over the poles
\eqn\etendoubleprime{
|\ell_7\rangle =|a\rangle +x_{-}|b\rangle, \quad
|\ell_8\rangle =|a\rangle +x_{+}|b\rangle.}

Thus, the bubble coefficient $c_{2:2;1}$ is represented by the sum
\eqn\eeighteen{
c_{2:2;1}=c_{2:2;1}^{(1)}+c_{2:2;1}^{(2)}+c_{2:2;1}^{(3)},}
where $c_{2:2;1}^{(1)}$, $c_{2:2;1}^{(2)}$ and $c_{2:2;1}^{(3)}$
are given by \eone, \etwo\ and~\enine\
respectively. The three-mass triangle coefficient $c_{3:2:2;1}^{3m}$ is
given by \esixteen.

\subsec{The amplitude $A(1^-,2^-,3^-,4^+,\ldots,n^+)$}

The next-to-MHV $n$-point amplitude with all negative-helicity gluons appearing consecutively has been computed in \BidderRI.
We perform the calculation here as well, as an illustration of our procedure.  The solution will emerge in a different form.

From the viewpoint of our discussion in Section 2, there
are no box and three-mass triangle contributions, hence no
one-mass and two-mass triangle contributions. All we have
are bubble contributions which can be calculated by double cuts.
There are two nonvanishing double cuts, which we denote by I and II.
\eqn\Ancuts{\eqalign{
{\rm I}  = & \int d\mu~ A_L(\ell_1,k+1,..., n,1,2,\ell_2) A_R(\ell_2, 3,4,...,k,\ell_1)
~~~~4\leq k\leq n-1\cr
{\rm II}  = &\int d\mu~ A_L(\ell_1,2,3,...,\W k,\ell_2) A_R(\ell_2,\W k+1,...,1,\ell_1)
~~~~~4\leq \W k\leq n-1
}}
However, these two kinds of cuts are mapped to each other under
the permutation $P_\a: (n-i+4)\leftrightarrow  i$
and the identification $\W k=n-k+3$.

Now we focus on the case ${\rm I}$. To do this we need to
know the tree-level amplitudes with two fermions or complex
scalars, which we list in Appendix B.
First we have
\eqn\AnFS{\eqalign{
 & A_L(\ell_1^+,(k+1)^+,..., n^+,1^-,2^-,\ell_2^-)\cr
 = &- \sum_{j=0}^{n-k-2} {  \braket{n-1-j~n-j}
\braket{1|K_{n-j}^{[j+2]}K_{n-j}^{[j+3]}|\ell_2}^3
\over \braket{\ell_2~\ell_1} \braket{\ell_1~k+1}\braket{k+1~k+2} \cdots
\braket{n~1}  t_{n-j}^{[j+2]}t_{n-j}^{[j+3]} \gb{n-j|K_{n-j}^{[j+2]}|2}
\gb{n-1-j|K_{n-j}^{[j+2]}|2}} \cr
 & \times\left( -{\braket{1|K_{n-j}^{[j+2]}
K_{n-j}^{[j+3]}|\ell_1}  \over \braket{1|K_{n-j}^{[j+2]}
K_{n-j}^{[j+3]}|\ell_2}} \right)^a \cr
 & -{  \braket{\ell_1~k+1}
\braket{1|K_{k+1}^{[n+2-(k+1)]} K_{k+1}^{[n+3-(k+1)]}  |\ell_2}^3
\over \braket{\ell_2~\ell_1} \braket{\ell_1~k+1}\braket{k+1~k+2}...
\braket{n|1} t_{k+1}^{[n+2-(k+1)]} t_{k+1}^{[n+3-(k+1)]}
\gb{k+1| K_{k+1}^{[n+2-(k+1)]} |2}
} \cr
 &\times
{1\over\gb{\ell_1| K_{k+1}^{[n+2-(k+1)]} |2}}
 \left(-{\braket{1|K_{k+1}^{[n+2-(k+1)]} K_{k+1}^{[n+3-(k+1)]}|\ell_1}
\over \braket{1|K_{k+1}^{[n+2-(k+1)]} K_{k+1}^{[n+3-(k+1)]}|\ell_2}} \right)^a
}}
where the last term is the special case with $j=n-k-1$. In this
formula, $a=0,1,2$ for $\ell_1, \ell_2$ to be gluons, fermions and
complex scalars. Using \AnFS\ and MHV-amplitude of $A_R$ we
get the integrand of cut ${\rm I}$ as
\eqn\Ancutsint{\eqalign{
{\rm I}  = & -\int d\mu{\braket{k~k+1}\braket{1~2} \braket{2~3}\over
\prod_{i=1}^n \braket{ i~i+1}}
{ \braket{\ell_1~3}\over  \braket{\ell_1~k+1}
\braket{k~\ell_1}} \cr
& \times \sum_{j=0}^{n-k-1} { \braket{n-1-j~n-j}
\braket{1|K_{n-j}^{[j+2]}K_{n-j}^{[j+3]}|3}^2
\braket{1|K_{n-j}^{[j+2]}K_{n-j}^{[j+3]}|\ell_1}
\over t_{n-j}^{[j+2]}t_{n-j}^{[j+3]} \gb{n-j|K_{n-j}^{[j+2]}|2}
\gb{n-1-j|K_{n-j}^{[j+2]}|2} }
}}

Notice that in the above integration, only the holomorphic
part $\ket{\ell_1}$ appears in the integrand, so the integration
is very easy to do, similar to the example $A(1^-,2^-,3^-,4^+, 5^+, 6^+)$.
We separate the integral into the cases where $j\neq n-k-1$ and the
special case of $j=n-k-1$. For the cases where $j\neq n-k-1$ there are
three potential poles:  $\ket{\ell_1} =  K_{3}^{[k-2]}|\eta]$,
$\ket{\ell_1} = \ket{k+1}$ and $\ket{\ell_1} = \ket{k}$. By choosing
$|\eta]= |K_{3}^{[k-2]}|3\rangle$, we get rid of one of them.
Similarly, for the case $j=n-k-1$, we take
$|\eta]= K_{3}^{[k-2]}|3\rangle$ to get rid of one pole and leave only
two poles: $\ket{\ell_1} = \ket{k}$ and $\ket{\ell_1}=|K_{2}^{[k-1]}|2]$.
Adding all these pieces together, we find that the coefficient from
cut I is
\eqn\AncoeffA{\eqalign{
c_{2:k-2;3}  = &  {\braket{1~2} \braket{2~3}\over
\prod_{i=1}^n \braket{ i~i+1}} \sum_{j=0}^{n-k-2}
{ \braket{n-1-j~n-j}
\braket{1|K_{n-j}^{[j+2]}K_{n-j}^{[j+3]}|3}^2 \over
t_{n-j}^{[j+2]}t_{n-j}^{[j+3]} \gb{n-j|K_{n-j}^{[j+2]}|2}
\gb{n-1-j|K_{n-j}^{[j+2]}|2}} \cr
 & \times \left( {\gb{3| K_{3}^{[k-2]}|k+1}
\braket{k+1| K_{n-j}^{[j+3]}K_{n-j}^{[j+2]}|1} \over
\gb{k+1|K_{3}^{[k-2]}|k+1}}
-{\gb{3| K_{3}^{[k-2]}|k} \braket{k| K_{n-j}^{[j+3]}K_{n-j}^{[j+2]}|1}
\over \gb{k|K_{3}^{[k-2]}|k}}\right) \cr
 & + {\braket{1~2} \braket{2~3}\over
\prod_{i=1}^n \braket{ i~i+1}}{\braket{k~k+1}\braket{1|K_{2}^{[k-1]}K_{3}^{[k-2]}|3}^2\over t_{2}^{[k-1]} t_{3}^{[k-2]} \gb{k+1|K_{2}^{[k-1]}|2}\gb{k|K_{2}^{[k-1]}|2}}  \cr
 & \times \left( -{ \gb{3|K_{3}^{[k-2]}|k} \braket{k| K_{3}^{[k-2]}
K_{2}^{[k-1]} |1}
 \over
\gb{k|K_{3}^{[k-2]}|k} }
 -{  \braket{2|K_{2}^{[k-1]}K_{3}^{[k-2]}|3}\gb{1|K_{3}^{[k-2]}|2} \over
 \gb{2|K_{3}^{[k-2]}|2}
}\right). \cr
}}
By symmetry we read out the coefficient from cut ${\rm II}$ as
\eqn\AncoeffB{\eqalign{
 c_{2:k-1;2}
 = & - {\braket{1~2} \braket{2~3}\over
\prod_{i=1}^n \braket{ i~i+1}} \sum_{j=0}^{n-k-2}
{\braket{j+5~j+4} \braket{3|K_{3}^{[j+2]}K_{2}^{[j+3]}|1}^2
\over t_{3}^{[j+2]}t_{2}^{[j+3]} \gb{j+4|K_{3}^{[j+2]}|2}
\gb{j+5|K_{3}^{[j+2]}|2}} \cr
 & \times
 \left( {\gb{1| K_{n-k+4}^{[k-2]}|n-k+3} \braket{n-k+3|K_{2}^{[j+3]}
 K_{3}^{[j+2]}|3}\over \gb{n-k+3| K_{n-k+4}^{[k-2]}|n-k+3}} \right.
 \cr  & \left. -
{\gb{1| K_{n-k+4}^{[k-2]}|n-k+4} \braket{n-k+4|K_{2}^{[j+3]}
 K_{3}^{[j+2]}|3}\over \gb{n-k+4| K_{n-k+4}^{[k-2]}|n-k+4}}\right)
\cr
&  + {\braket{1~2} \braket{2~3}\over
\prod_{i=1}^n \braket{ i~i+1}} { \braket{n-k+4~~n-k+3}
\braket{3|K_{n-k+4}^{[k-1]}K_{n-k+4}^{[k-2]}|1}^2 \over
t_{n-k+4}^{[k-1]}t_{n-k+4}^{[k-2]}   \gb{n-k+3|K_{n-k+4}^{[k-1]}|2}
 \gb{n-k+4|K_{n-k+4}^{[k-1]}|2}} \cr
 & \times \left( {\gb{1|K_{n-k+4}^{[k-2]}|n-k+4} \braket{n-k+4| K_{n-k+4}^{[k-2]}K_{n-k+4}^{[k-1]}|3} \over \gb{n-k+4|K_{n-k+4}^{[k-2]} |k-k+4}}
\right. \cr
 & \left. +{ \braket{2|K_{n-k+4}^{[k-1]} K_{n-k+4}^{[k-2]}|1}
\gb{3|K_{n-k+4}^{[k-2]}|2} \over \gb{2|K_{n-k+4}^{[k-2]} |2}}\right).
}}

In \BidderRI, the amplitude was decomposed in terms of functions ${\rm K}_0$ and ${\rm L}_0$, given here in \quaint, where the function ${\rm K}_0$ is proportional to the bubble integral and the function ${\rm L}_0$ is related to the Feynman parameter integral for a two-mass triangle integral.  By the identity \sotoo, it is possible to convert their expression to an expansion in bubble integrals only, as we have done here.


\newsec{Summary Of Results For Next-To-MHV Six-Gluon Amplitudes}

In this section we collect our results for the next-to-MHV six-gluon amplitudes for the reader's
convenience.

We define the following functions:
\eqn\Rfun{\eqalign{
R_1 (a_j,b_j,c_j) & =
{   (2 a_2 b_1 c_1-a_1 b_2 c_1-a_1 b_1 c_2+
2 a_0 b_2 c_2) \over 2( a_2 b_1^2 - a_1 b_1 b_2+ a_0 b_2^2)}
\cr
R_2(a_j,b_j,c_j) & = {(a_2b_2c_1 +a_2b_1c_2 -a_1b_2c_2) \over b_2(2a_2b_1-a_1b_2)}}}
%

\subsec{$A(1^-,2^-,3^-,4^+,5^+,6^+)$}
The amplitude is given by
\eqn\adjcentTotalone{\eqalign{
A(1^-,2^-,3^-,4^+,5^+,6^+) = &{r_\Gamma (\mu^2)^{\epsilon} \over
 (4\pi)^{2-\epsilon}}
\left( c_{2:3;6} I_{2:3;6} +c_{2:3;2}I_{2:3;2}+
c_{2:2;3}I_{2:2;3}+ c_{2:2;6}I_{2:2;6}\right)
}}
with $r_\Gamma = {\Gamma(1+\epsilon)\Gamma^2(1-\epsilon) \over \Gamma(1-2\epsilon)}$ and
\eqn\adjcentTotaltwo{\eqalign{
c_{2:3;6}
 = & -{ \gb{3|P_{612}|6}^2 \over [6~1][1~2] \vev{3~4}
\vev{4~5} \gb{5|P_{612}|2} P_{612}^2} \left(
{ \gb{3|P_5|6} P_{612}^2 \over \gb{5|P_{612}|5}}
+{\gb{3|P_{612} P_2 P_{612} |6} \over \gb{2|P_{612}|2}}\right)\cr
c_{2:3;2}
= & -{  \gb{1|P_{561}|4}^2 \over \vev{5~6}\vev{6~1}
[2~3][3~4] \gb{5|P_{561}|2} P_{561}^2} \left(
{ \gb{1|P_2|4}P_{561}^2 \over \gb{2|P_{561}|2}}+
{\gb{1|P_{561}P_5 P_{561}|4} \over \gb{5|P_{561}|5}} \right)\cr
c_{2:2;3}
 = & { [4~5] \gb{5|P_{345}|6} \gb{3|P_{345}|6}^2
\over \vev{4~5} [6~1][1~2] \gb{5|P_{345}|5} \gb{5|P_{345}|2} P_{345}^2}
+{ \vev{2~3} \gb{1|P_{234}|2} \gb{1|P_{234}|4}^2
\over [2~3] \vev{5~6}\vev{6~1} \gb{2|P_{234}|2} \gb{5|P_{234}|2}
P_{234}^2}\cr
c_{2:2;6}
= & { [5~6] \gb{5|P_{561}|4} \gb{1|P_{561}|4}^2 \over
\vev{5~6} [2~3] [3~4] \gb{5|P_{561}|2} \gb{5|P_{561}|5} P_{561}^2}
+{ \vev{1~2} \gb{3|P_{612}|2} \gb{3|P_{612}|6}^2 \over
[1~2] \vev{3~4}\vev{4~5} \gb{5|P_{612}|5} \gb{2|P_{612}|2} P_{612}^2}
}}

\subsec{$A(1^-,2^-,3^+,4^-,5^+,6^+)$}
The amplitude is given by
\eqn\secondTotalone{\eqalign{
A  (1^-,2^-,3^+,4^-,5^+,6^+) = & {r_\Gamma (\mu^2)^{\epsilon} \over
 (4\pi)^{2-\epsilon}}
\left( c^{2m~h}_{4:2;2}I^{2m~h}_{4F:2;2}+c^{2m~h}_{4:2;4} I^{2m~h}_{4F:2;4}
+ c^{2m~h}_{4:2;6} I^{2m~h}_{4F:2;6} +c^{1m}_{4;5}I^{1m}_{4F;5}
+c^{1m}_{4;6} I^{1m}_{4F;6} \right. \cr &
 +c^{3m}_{3:2:2;2}I^{3m}_{3:2:2;2}+
c_{2:3;2}I_{2:3;2}+c_{2:3;6}I_{2:3;6}+c_{2:3;1}I_{2:3;1} \cr & \left.
+c_{2:2;2}I_{2:2;2}+c_{2:2;3}I_{2:2;3}+c_{2:2;4}I_{2:2;4}
+c_{2:2;6}I_{2:2;6}\right)
}}
with
\eqn\secondTotaltwo{\eqalign{
c^{2m~h}_{4:2;2}  = &-{P_{61}^2 P_{456}^2\over 2}
 { \gb{4|P_{456}|3}^2  \braket{4~6}[3~1]
\over \gb{6|P_{456}|1}^2 \braket{4~5}\braket{5~6}[1~2][2~3]}\cr
c^{2m~h}_{4:2;4}  = & -{P_{23}^2 P_{612}^2 \over 2}
{ [2~6] \gb{4|P_{612}|2} \gb{4|P_{612}|6}^2
\over \braket{4~5} [6~1][1~2] \gb{5|P_{612}|2}\gb{3|P_{612}|2}^2} \cr
c^{2m~h}_{4:2;6}
  = & -{P_{45}^2 P_{561}^2 \over 2}
{\braket{1~5}\gb{1|P_{561}|3}^2 \gb{5|P_{561}|3}
 \over [2~3]\braket{5~6}\braket{6~1}
\gb{5|P_{561}|2} \gb{5|P_{561}|4}^2} \cr
c^{1m}_{4;5}  = & -{P_{23}^2 P_{34}^2\over 2}
{  \gb{1|P_{234}|2} \gb{1|P_{234}|3}^2
\over \braket{5~6}\braket{6~1} [4~2]^2 \gb{5|P_{234}|2} P_{234}^2} \cr
c^{1m}_{4;6}  = & -{P_{34}^2 P_{45}^2 \over 2}
{ \gb{5|P_{345}|6} \gb{4|P_{345}|6}^2
\over [6~1][1~2] \braket{3~5}^2 \gb{5|P_{345}|2} P_{345}^2}
}}
\eqn\secondTotalthree{\eqalign{
& c^{3m}_{3:2:2;2}  = {1\over [4~5]\braket{6~1}\braket{2~3}} \times
\cr
& ~~~ \sum_{i=1,2,3,6,7}\left[ \braket{\ell~\ell_i}{  g(\ell) \gb{\ell|Q|3}
R_1(a_j,b_j,c_j)\over \braket{\ell~\W \eta}
\gb{\ell|Q|2} \braket{\ell|P_{23} Q|\ell}} \left(
\gb{1|6|5}\braket{\ell|P_{23} Q|\ell}+ P_{23}^2
\braket{1~\ell}\gb{\ell|Q|5} \right)\right]_{\ell\to \ell_i}
\cr
& ~~~ + {1\over [4~5]\braket{6~1}\braket{2~3}} \times 
\cr 
& ~~~ \sum_{i=4,5}\left[ \braket{\ell~\ell_i}{  g(\ell) \gb{\ell|Q|3}
R_2(a_j,b_j,c_j)\over \braket{\ell~\W \eta}
\gb{\ell|Q|2} \braket{\ell|P_{23} Q|\ell}} \left(
\gb{1|6|5}\braket{\ell|P_{23} Q|\ell}+ P_{23}^2
\braket{1~\ell}\gb{\ell|Q|5} \right)\right]_{\ell\to \ell_i}.
}}
\eqn\secondTotalfour{\eqalign{
c_{2:3;2}
  = & -{ \gb{1|P_{234}|3}^2 \over [2~3][3~4]\braket{5~6}
\braket{6~1} P_{234}^2} \left( { \gb{1|6|5} \gb{5|P_{561}|3}^2 \over
\gb{5|P_{561}|2}
\gb{5|P_{561}|4} \gb{5|P_{561}|5}} \right. \cr
&  \left. + { \braket{1~2} [2~3]^2 P_{561}^2
\over  [2~4]\gb{5|P_{561}|2} \gb{2|P_{561}|2}}
 + { \braket{1~4} [3~4]^2 P_{561}^2\over [4~2]\gb{5|P_{561}|4}
\gb{4|P_{561}|4}} \right) \cr
c_{2:3;6}
  = & -{  \gb{4|P_{612}|6}^2 \over [6~1][1~2] \braket{3~4}
\braket{4~5} P_{612}^2} \left( {\gb{2|1|6} \gb{4|P_{612}|2}^2
\over \gb{5|P_{612}|2} \gb{3|P_{612}|2} \gb{2|P_{612}|2}}
\right. \cr
 & \left. + { [5~6] \braket{4~5}^2 P_{612}^2\over \braket{3~5}
\gb{5|P_{612}|2} \gb{5|P_{612}|5}} +{ [6~3]\braket{3~4}^2
P_{612}^2\over \braket{3~5} \gb{3|P_{612}|2} \gb{3|P_{612}|3}}
\right) \cr
c_{2:3;1}
 = &{ \gb{4|P_{456}|3}^2  \over [1~2][2~3]
\braket{4~5}\braket{5~6} }
\left[ { [6~3] \braket{6~4} \over \gb{6|P_{123}|6}
\gb{6|P_{123}|1}} -{ \gb{1|2|3} \gb{4|P_{123}|1}\over  \gb{1|P_{123}|1}
\gb{6|P_{123}|1} P_{123}^2} \right]
}}
\eqn\secondTotalfive{\eqalign{
c_{2:2;2} = & {\braket{2~4}^2 [5~6]^4 \over [5~6][6~1]
P_{561}^2} {\gb{4|P_{561}|5} \over \gb{4|P_{561}|1}}
{\gb{5|P_{561}|3} \over \gb{3|P_{561}|5}} {1\over
\gb{5|P_{561} P_{23} P_{561}|5}}\cr
+ & {\gb{4|P_{456}|3}^2 \over [2~3] \braket{4~5}\braket{5~6}
P_{456}^2 \gb{4|P_{456}|1}} \left( -{[3~2] \braket{2~1}
\gb{4|P_{456}|1}^2 \over [2~1]\gb{6|P_{456}|1} \gb{1|P_{23}|1}}
\right. \cr
 & \left. + {  \braket{4~6}^2 [2~3] \gb{2|P_{456}|6}
(P_{456}^2)^2 \over \gb{6|P_{456}|2} \gb{6|P_{456}|1}
\gb{6|P_{456} P_{23} P_{456}|6}} \right)\cr
- & {1\over [4~5]\braket{6~1}}\sum_{i=1}^5
\lim_{\ell\to \ell_i} \left[ \braket{\ell~\ell_i}{  g(\ell)
\braket{1~\ell} \gb{\ell|P_{23}|5} [3~\ell]\over
\braket{\ell~3}\braket{\ell|P_{23} Q|\ell}\gb{\ell|P_{23}|\ell}}\right]\cr
c_{2:2;3}
  = & { [3~5] \gb{5|P_{612}|6} \gb{4|P_{612}|6}^2 \over [6~1][1~2]
\braket{3~5} \gb{5|P_{612}|2} \gb{5|P_{612}|5} P_{612}^2}
+ { \vev{2~4}\gb{1|P_{561}|2} \gb{1|P_{561}|3}^2 \over \braket{5~6}
\braket{6~1}[2~4] \gb{5|P_{561}|2} \gb{2|P_{561}|2} P_{561}^2}\cr
c_{2:2;4}
  = & P_\a (c_{2:2;2}) \cr
c_{2:2;6}
  = & A_{\rm tree}-c_{2:2;2}-c_{2:2;3}-c_{2:2;4}-c_{2:3;2}-c_{2:3;6}-c_{2:3;1}
}}
The symmetric
action is $P_\a: ~i\leftrightarrow 7-i$ plus conjugation, i.e,
$\braket{~~}\leftrightarrow [~~]$.
For \secondTotalthree\ and \secondTotalfive, we have defined
\eqn\intpara{\eqalign{
a_0  = & (Q-P_{23})^2,~~~~~a_1= 2 P_{23} \cdot (Q-P_{23}),~~~~~
a_2= P_{23}^2 \cr
b_1  = & \gb{\ell| (Q-P_{23})|\ell},~~~~b_2=\gb{\ell|P_{23}|\ell}
~~~~c_1=\gb{\W \eta|(Q-P_{23})|\ell},~~~~c_2=\gb{\W \eta|P_{23}|\ell}
}}
\eqn\sevvpoles{\eqalign{
\ket{\ell_1}= & |P_{23}|4],~~~\ket{\ell_2}=|P_{61}|5],~~~~\ket{\ell_3}=
|P_{23} P_{45}|6\rangle \cr
\ket{\ell_4}= & \ket{a}+ x_+\ket{b},~~~~~
\ket{\ell_5}=  \ket{a}+ x_-\ket{b},~~~~
\ket{\ell_6}= |Q|2],~~~~\ket{\ell_7}=\ket{\W \eta}}}
\eqn\unkownx{ x_\pm={ -(\braket{a|P_{23} Q|b}+\braket{b|P_{23} Q|a})\pm
 \braket{a~b} \sqrt{\Delta_{3m}}\over 2
\braket{b|P_{23}Q|b}}}
where
\eqn\unkownPQ{\eqalign{
\Delta_{3m} &  =   (P_{61}^2)^2+(P_{45}^2
)^2+(P_{23}^2)^2-2 P_{61}^2 P_{23}^2 -2 P_{23}^2 P_{45}^2 -2
P_{45}^2 P_{61}^2 \cr
g(\ell) & =
-{\gb{\ell|P_{23}|5} \over \gb{\ell|P_{23}|4}}
{ ( \gb{1|6|5}\braket{2~\ell}+\gb{2|3|5}\braket{1~\ell})^2
\over \gb{\ell|P_{61}|5} \braket{\ell|P_{23} P_{45}|6}}\cr
Q  & =  {1 \over P_{23}^2}(P_{61}^2 P_{23}+   P_{23}^2 P_{61})
}}%
Here $\ket{a},\ket{b}$ and $\ket{\W \eta}$ are arbitrary.

\subsec{$A(1^-,2^+,3^-,4^+,5^-,6^+)$}

This helicity configuration has the largest symmetry, namely a
$\QZ_6$ generated by $P_\a: i\to i+1$ plus conjugation, so
we have grouped everything into orbits. The amplitude
is given by
\eqn\ThirdTotalone{\eqalign{
A  (1^-,2^+,3^-,4^+,5^-,6^+) = & {r_\Gamma (\mu^2)^{\epsilon} \over
 (4\pi)^{2-\epsilon}}
\left( \sum_{i=1}^6 c^{2m~h}_{4:2;i}I^{2m~h}_{4F:2;i}
+\sum_{i=1}^6 c^{1m}_{4;i} I^{1m}_{4F;i} +\sum_{i=1}^2
c^{3m}_{3:2:2;i}I^{3m}_{3:2:2;i} \right. \cr & \left.
+ \sum_{i=1}^3 c_{2:3;i} I_{2:3;i}
+ \sum_{i=1}^6 c_{2:2;i} I_{2:2;i}  \right)
}}
with
\eqn\ThirdTotaltwo{\eqalign{
c^{2m~h}_{4:2;1} =&  { P_{345}^2 P_{56}^2 \over 2}
{ \gb{1|P_{345}|4}^2 \over
\braket{1~2}[3~4] \gb{6|P_{345}|5}^2}
{\gb{1|P_{345}|5}\gb{6|P_{345}|4} \over
\gb{2|P_{345}|5}\gb{6|P_{345}|3}} \cr
c^{1m}_{4;1}  =&  -{P_{45}^2 P_{56}^2\over 2}
 { \gb{5|P_{456}|2}^2
\over [1~2][2~3] \braket{4~6}^2 P_{456}^2} { \gb{4|P_{456}|2}
\gb{6|P_{456}|2}\over \gb{4|P_{456}|1} \gb{6|P_{456}|3}} \cr
c_{3:2:2;1}^{3m} =& {1 \over \braket{1~2} [3~4] \braket{5~6}}
\cr & \times
\sum_{i=1,2,3,4,7,8} \left[\braket{\ell \ell_i} 
{g(\ell)\gb{\ell|Q|2}R_1(a_j, b_j, c_j) 
\over \braket{\ell~\tilde{\eta}} \gb{\ell|Q|1} \braket{\ell|P_{12}Q|\ell}}
(\gb{5|6|4}\braket{\ell|P_{12}Q|\ell}+P_{12}^2\braket{5~\ell} \gb{\ell|Q|4})
\right]_{\ell \to \ell_i} \cr +&
{1 \over \braket{1~2} [3~4] \braket{5~6}}
\sum_{i=5,6} \left[\braket{\ell \ell_i} 
{g(\ell)\gb{\ell|Q|2}R_2(a_j, b_j, c_j) 
\over \braket{\ell~\tilde{\eta}} \gb{\ell|Q|1} \braket{\ell|P_{12}Q|\ell}}
(\gb{5|6|4}\braket{\ell|P_{12}Q|\ell}+P_{12}^2\braket{5~\ell} \gb{\ell|Q|4})
\right]_{\ell \to \ell_i},
\cr
c_{2:3;1}
  =&  -{ \gb{5|P_{123}|2}^2 \over [1~2][2~3]\braket{4~5}\braket{5~6}
P_{123}^2}\left( { \gb{4|P_{123}|2}^2 \braket{4~5}\braket{5~6}[4~6]
\over \gb{4|P_{123}|1}\gb{4|P_{123}|3}\gb{4|P_{123}|4}\braket{4~6}}
\right. \cr
&  \left.  -{ \braket{1~5}[1~2]^2 P_{123}^2 \gb{5|P_{123}|1}
\over [1~3] \gb{4|P_{123}|1}\gb{6|P_{123}|1}\gb{1|P_{123}|1}}
+ {\gb{5|P_{123}|3}\over \gb{6|P_{123}|3}}
{ \braket{5~3} [2~3]^2 P_{123}^2 \over [3~1]
\gb{4|P_{123}|3} \gb{3|P_{123}|3}}\right. \cr
 & \left. +  { \braket{5~6}\over \gb{6|P_{123}|3}}{ \braket{5~4}[4~6]
\gb{6|P_{123}|2}^2 \over \braket{6~4} \gb{6|P_{123}|1}
\gb{6|P_{123}|6}} \right)
\cr
c_{2:2;1}
  =& {[4~6]^4\braket{1~3}^2 \over [4~5][5~6]\gb{3|P_{456}|6}P^2_{456}}
{\gb{3|P_{456}|4} \gb{4|P_{456}|2} \over \gb{2|P_{456}|4}
\gb{4|P_{456}P_{12}P_{456}|4}}\cr
& +  {\braket{3~5}^4 [2~6]^2 \over \braket{3~4}
\braket{4~5}\gb{3|P_{345}|6}P^2_{345}}
{\gb{1|P_{345}|5} \gb{5|P_{345}|6} \over \gb{5|P_{345}|1}
\gb{5|P_{345}P_{12}P_{345}|5}}\cr
& -{1 \over [3~4]\braket{5~6}} \sum_{i=1}^6
\lim_{\ell \to \ell_{i}} \left[\braket{\ell~\ell_i}
{g(\ell) \braket{5~\ell} \gb{\ell|P_{12}|4} [2~\ell] \over
\braket{\ell~2} \braket{\ell|P_{12}Q|\ell} \gb{\ell|P_{12}|\ell}}
\right]
}}
and the other coefficients can be obtained by the symmetric action of $P_\a$.
Here one needs
\eqn\etenvv{\eqalign{
&|\ell_1\rangle =|P_{56}|4], \quad
|\ell_2\rangle =|P_{12}|3], \quad
|\ell_3\rangle =|6 \rangle, \quad
|\ell_4\rangle =|P_{12}P_{34}|5 \rangle, \cr
&|\ell_5\rangle =|a\rangle +x_{-}|b\rangle, \quad
|\ell_6\rangle =|a\rangle +x_{+}|b\rangle, \quad
|\ell_7\rangle =|\tilde{\eta} \rangle, \quad
|\ell_8\rangle =|Q|1].}}
\eqn\eesixteenone{\eqalign{
& a_0=(Q-P_{12})^2, \quad a_1=2P_{12}\cdot (Q-P_{12}), \quad a_2=P_{12}^2, \cr
& b_1 =\gb{\ell|Q-P_{12}|\ell}, \quad b_2 =\gb{\ell|P_{12}|\ell}, \quad
c_1 = \gb{\tilde{\eta}|(Q-P_{12})|\ell}, \quad c_2=\gb{\tilde{\eta}|P_{12}|\ell}}}
\eqn\eseven{\eqalign{
g(\ell) &= -{\braket{\ell~5}\over \braket{\ell~6}}
{\gb{\ell|P_{12}|4} \over \braket{\ell|P_{12}P_{34}|5}}
{(\gb{5|6|4}\braket{1~\ell}+\gb{1|2|4}\braket{5~\ell})^2
\over \gb{\ell|P_{56}|4} \gb{\ell|P_{12}|3}} \cr
 Q &= {1 \over P_{12}^2}(P_{56}^2P_{12}+P_{12}^2P_{56}).}}
\eqn\unkownxtwo{ x_\pm={ -(\braket{a|P_{12} Q|b}+\braket{b|P_{12} Q|a})\pm
 \braket{a~b} \sqrt{\Delta_{3m}}\over 2
\braket{b|P_{12}Q|b}}}
where
\eqn\unkownPQtwo{\eqalign{
\Delta_{3m} &   =   (P_{12}^2)^2+(P_{34}^2
)^2+(P_{56}^2)^2-2 P_{12}^2 P_{34}^2 -2 P_{34}^2 P_{56}^2 -2
P_{56}^2 P_{12}^2
}}
and $\ket{a}, \ket{b}, \ket{\W \eta}$ can be chosen arbitrarily.

\bigskip
\bigskip
\centerline{\bf Acknowledgments}

We thank R. K. Ellis, P. Mastrolia, R. Roiban, M. Spradlin, P. Svr\v{c}ek and A. Volovich
for helpful comments and discussions, and L. Dixon for identifying a typo in the previous versions of formulas for $c^{3m}_{3:2:2;2}$.
R.B., E.B. and B.F. were supported by NSF grant PHY-0070928. F.C. was
supported in part by the Martin A. and Helen Chooljian Membership
at the Institute for Advanced Study and by DOE grant
DE-FG02-90ER40542.

\appendix{A}{Scalar Integral Functions}

In this appendix we list the explicit results for the scalar integrals derived in the reduction procedure of \BernKR.  The expressions here are taken from \refs{\BernCG,\DennerQQ}.

The dimensional regularization parameter is $\epsilon=(4-D)/2$.  The constant $r_\Gamma$ is defined by
\eqn\rgamma{r_\Gamma = {\Gamma(1+\epsilon)\Gamma^2(1-\epsilon) \over \Gamma(1-2\epsilon)}
}

\ifig\ponhug{Scalar bubble and triangle integrals. (a) One-mass triangle $I^{1m}_{3;i}$. (b) Two-mass triangle $I^{2m}_{3:r;i}$.  (c) Three-mass triangle $I^{3m}_{3:r:r';i}$.  (d) Bubble $I_{2:r;i}$.}
{\epsfxsize=0.70\hsize\epsfbox{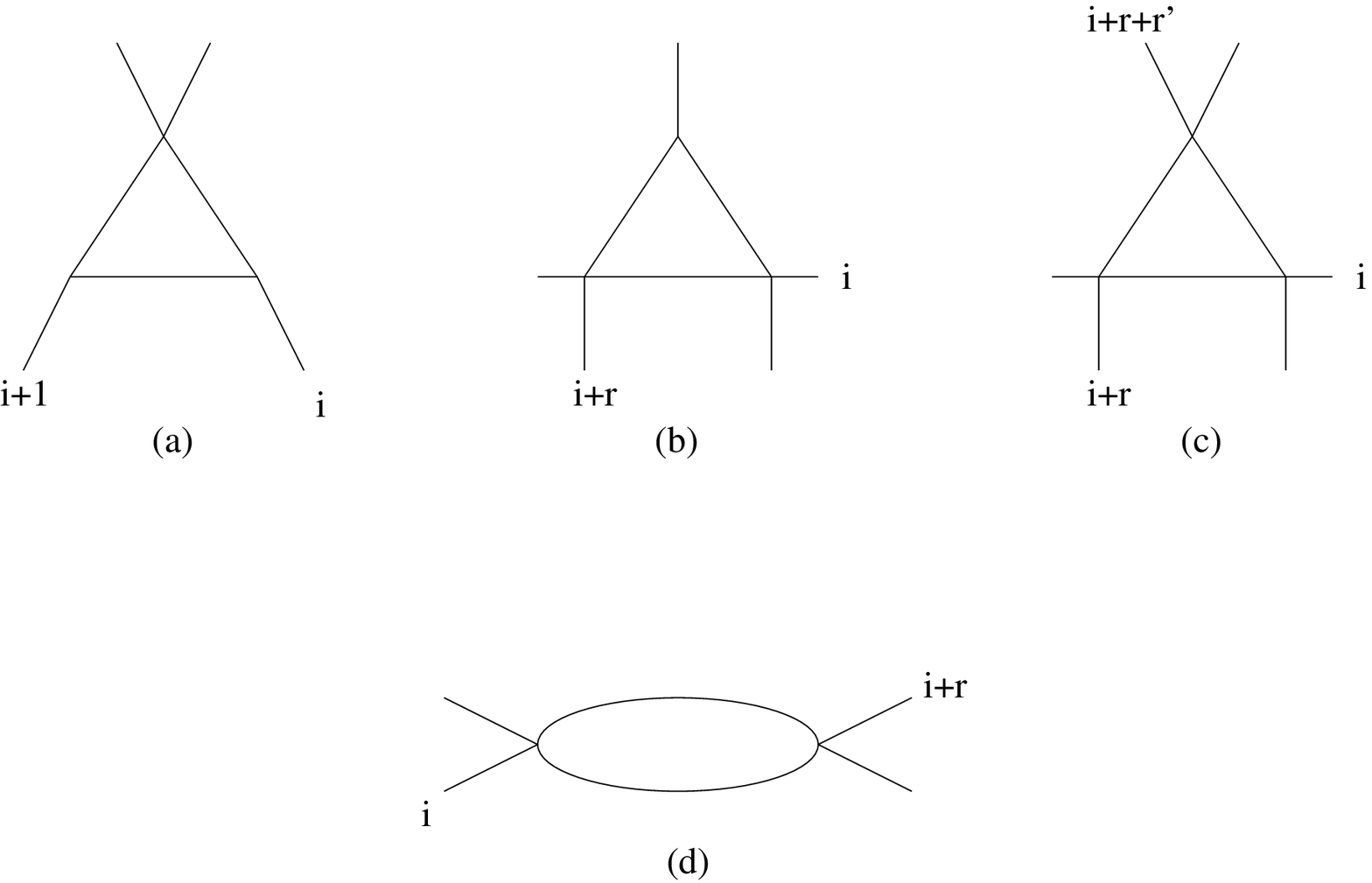}}

\ifig\kkmbp{Scalar box integrals.  (a) The outgoing external
momenta at each of the vertices are $K_1,K_2,K_3,K_4$, defined to
correspond to sums of the momenta of gluons in the exact
orientation shown. (b) One-mass $I^{1m}_{4;i}$. (c) Two-mass
``easy" $I^{2m~e}_{4:r;i}$. (d) Two-mass ``hard"
$I^{2m~h}_{4:r;i}$. (e) Three-mass $I^{3m}_{4:r:r';i}$. (f)
Four-mass $I^{4m}_{4:r:r':r";i}$. }
{\epsfxsize=0.70\hsize\epsfbox{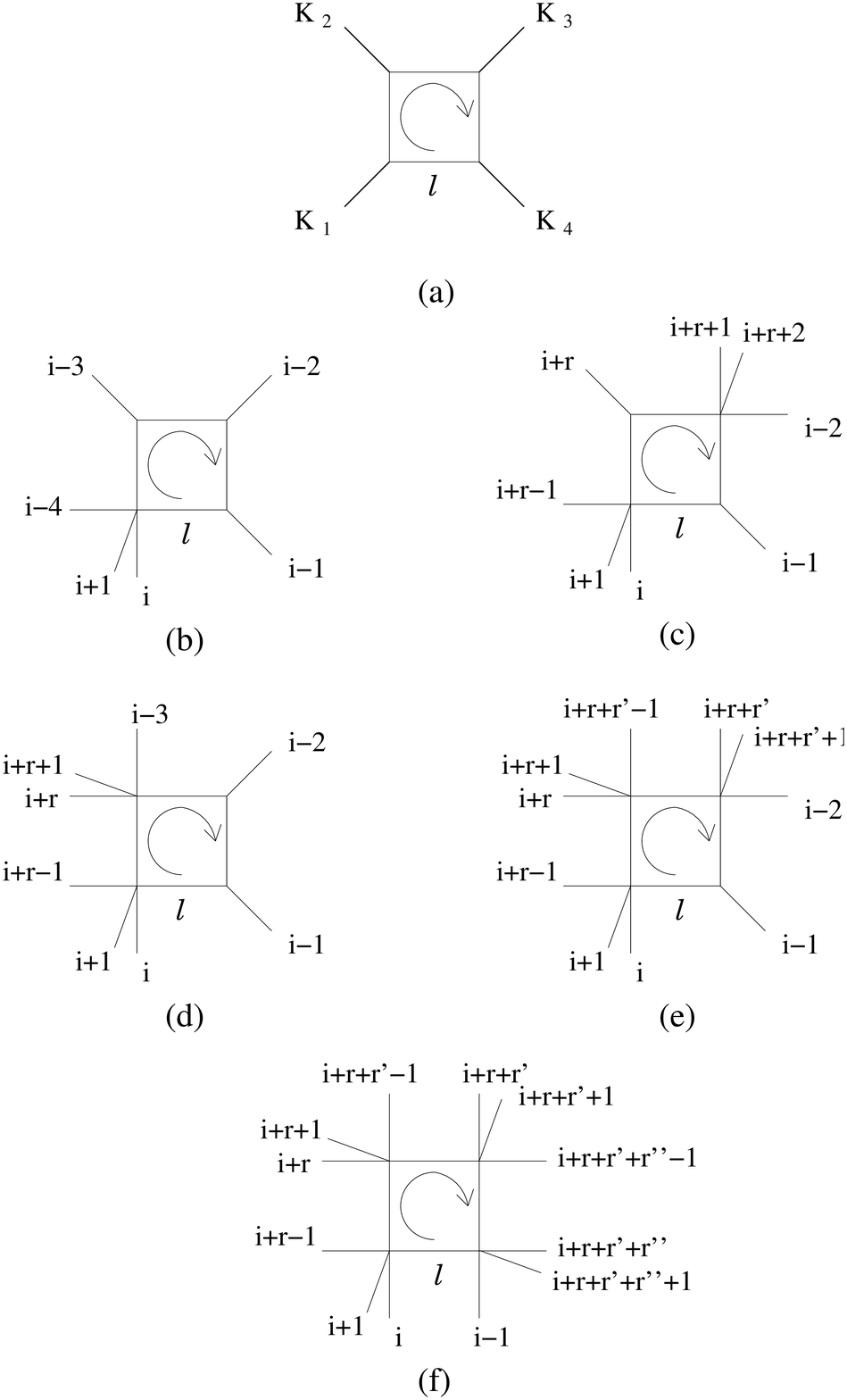}}

Scalar bubble integrals:

\eqn\tricho{I_2 = -i (4\pi)^{2-\epsilon} \int {d^{4-2\epsilon} p\over
(2\pi)^{4-2\epsilon}} {1\over p^2(p-K)^2}}
\eqn\mermith{
I_{2:r;i}=r_{\Gamma}
\left({1\over \epsilon} -\ln(-t_{i}^{[r]})+2\right) +\CO(\epsilon)}

Scalar triangle integrals:

\eqn\doryla{I_3  =  i (4\pi)^{2-\epsilon} \int {d^{4-2\epsilon} p\over
(2\pi)^{4-2\epsilon}} {1\over p^2(p-K_1)^2(p+K_3)^2}}

\eqn\mononch{\eqalign{
I^{1m}_{3;i}& = { r_{\Gamma} \over \epsilon^2} (-t_{i}^{[2]})^{-1-
\epsilon}\cr
I^{2m}_{3:r;i} & =  { r_{\Gamma} \over \epsilon^2} {
(-t_{i}^{[r]})^{-\epsilon} -(-t_{i+r}^{[n-r-1]})^{-\epsilon}
\over (-t_{i}^{[r]})-(-t_{i+r}^{[n-r-1]})} \cr
I^{3m}_{3:r:r';i} & =  {i \over \sqrt{\Delta_3}} \sum_{j=1}^3
\left[ {\rm Li}_2\left(-{1+i\delta_j \over 1-i \delta_j} \right)
-{\rm Li}_2\left(-{1-i\delta_j \over 1+i \delta_j} \right)
\right]+\CO(\epsilon)
}}

We have defined the following:
\eqn\triplonch{\eqalign{
\Delta_3 & =  -(K_1^2)^2-(K_2^2)^2-(K_2^2)^2+ 2K_1^2 K_2^2
+2 K_2^2 K_3^2 +2 K_3^2 K_1^2 \cr
\delta_1 & =  { t_{i}^{[r]}-t_{i+r}^{[r']}-t_{i+r+r'}^{[n-r-r']}
\over \sqrt{\Delta_3}} \cr
\delta_2 & =  { -t_{i}^{[r]}+t_{i+r}^{[r']}-t_{i+r+r'}^{[n-r-r']}
\over \sqrt{\Delta_3}} \cr
\delta_3 & =  { -t_{i}^{[r]}-t_{i+r}^{[r']}+t_{i+r+r'}^{[n-r-r']}
\over \sqrt{\Delta_3}}
}}

Scalar box integrals:

\eqn\ifm{ I_4^{4m}= -i (4 \pi)^{2-\epsilon}
\int {d^4\ell \over (2\pi)^{4-2\epsilon}} { 1\over (\ell^2+i\epsilon) (
(\ell-K_1)^2+i\epsilon) ( (\ell -K_1-K_2)^2+i\epsilon)
((\ell+K_4)^2+i\epsilon) }.}

We separate divergent terms from the finite pieces of interest, as
explained more fully in section 2.  We refer to the finite pieces $I_{4F}$ as
finite box integral functions.

\eqn\expi{\eqalign{ I^{1m}_{4;i} & =
{2r_{\Gamma} \over t_{i-3}^{[2]} t_{i-2}^{[2]}} ~{1\over \epsilon^2}\left[
(-t_{i-3}^{[2]})^{-\epsilon} +(-t_{i-2}^{[2]})^{-\epsilon}
-(-t_{i-3}^{[3]})^{-\epsilon} \right] +I^{1m}_{4F;i}.
 \cr I^{1m}_{4F;i} & = -{2r_{\Gamma} \over t_{i-3}^{[2]} t_{i-2}^{[2]}}
\left[  {\rm Li}_2 \left( 1-
{t_{i-3}^{[3]}\over t_{i-3}^{[2]}} \right) + {\rm Li}_2 \left( 1-
{t_{i-3}^{[3]}\over t_{i-2}^{[2]}} \right) + {1\over 2}\ln^2\left(
{t_{i-3}^{[2]}\over t_{i-2}^{[2]}} \right) + {\pi^2\over 6}+\CO(\epsilon)
\right].}}
\eqn\exop{\eqalign{ I^{2m~e}_{4:r;i} & =
{2r_{\Gamma} \over t_{i-1}^{[r+1]}t_{i}^{[r+1]}-t_{i-1}^{[r+2]}t_{i}^{[r]}}
 ~ {1\over
\epsilon^2}\left[ (-t_{i-1}^{[r+1]})^{-\epsilon}
+(-t_{i}^{[r+1]})^{-\epsilon} -(-t_{i}^{[r]})^{-\epsilon} -
(-t_{i-1}^{[r+2]})^{-\epsilon} \right] +I^{2m~e}_{4F:r;i}
\cr I^{2m~e}_{4F:r;i} &= -{2r_{\Gamma} \over t_{i-1}^{[r+1]}t_{i}^{[r+1]}-t_{i-1}^{[r+2]}t_{i}^{[r]}}
\left[ {\rm Li}_2 \left(
1- {t_{i}^{[r]}\over t_{i-1}^{[r+1]}} \right) + {\rm Li}_2 \left(
1- {t_{i}^{[r]}\over t_{i}^{[r+1]}} \right) +{\rm Li}_2 \left( 1-
{t_{i-1}^{[r+2]}\over t_{i-1}^{[r+1]}} \right)\right. \cr & ~~+ \left.{\rm Li}_2
\left( 1- {t_{i-1}^{[r+2]}\over t_{i}^{[r+1]}} \right) - {\rm
Li}_2 \left( 1- {t_{i}^{[r]}t_{i-1}^{[r+2]}\over
t_{i-1}^{[r+1]}t_{i}^{[r+1]}} \right) + {1\over 2}\ln^2\left(
{t_{i-1}^{[r+1]}\over t_{i}^{[r+1]}} \right)+\CO(\epsilon)\right].}}

\eqn\uxop{\eqalign{ I^{2m~h}_{4:r;i} &=
{2r_{\Gamma} \over t_{i-2}^{[2]}t_{i-1}^{[r+1]}}
~ {1\over
\epsilon^2}\left[ \half (-t_{i-2}^{[2]})^{-\epsilon}
+(-t_{i-1}^{[r+1]})^{-\epsilon} -\half (-t_{i}^{[r]})^{-\epsilon} -
\half (-t_{i-2}^{[r+2]})^{-\epsilon} \right] +I^{2m~h}_{4F:r;i} \cr
I^{2m~h}_{4F:r;i} &=
-{2r_{\Gamma} \over t_{i-2}^{[2]}t_{i-1}^{[r+1]}}
\left[ -\half \ln\left({t_{i-2}^{[2]} \over t_i^{[r]}}\right)
\ln\left({t_{i-2}^{[2]} \over t_{i-2}^{[r+2]}}\right)
+ {1\over 2}\ln^2 \left(
{t_{i-2}^{[2]}\over t_{i-1}^{[r+1]}} \right)\right. \cr &~~
\left. +  {\rm Li}_2
\left( 1- {t_{i}^{[r]}\over t_{i-1}^{[r+1]}} \right) + {\rm Li}_2
\left( 1- {t_{i-2}^{[r+2]}\over t_{i-1}^{[r+1]}} \right)+\CO(\epsilon)\right]. }}

\eqn\uliz{\eqalign{ I^{3m}_{4:r:r';i} &=
{2r_{\Gamma} \over t_{i-1}^{[r+1]}t_i^{[r+r']}-t_i^{[r]}t_{i-1}^{[r+r'+1]}} \cr
&
 ~ {1\over \epsilon^2}
\left[\half (-t_{i-1}^{[r+1]})^{-\epsilon}
+\half (-t_{i}^{[r+r']})^{-\epsilon} -\half (-t_{i}^{[r]})^{-\epsilon}
- \half (-t_{i-1}^{[r+r'+1]})^{-\epsilon}
\right]+  I^{3m}_{4F:r:r';i}.   \cr
I^{3m}_{4F:r:r';i} &=-{2r_{\Gamma} \over t_{i-1}^{[r+1]}t_i^{[r+r']}-t_i^{[r]}t_{i-1}^{[r+r'+1]}}
\left[
-\half \ln\left({t_i^{[r+r']} \over t_i^{[r]}}\right)
\ln\left({t_i^{[r+r']} \over t_{i+r}^{[r+r']}}\right)
-\half \ln\left({t_{i-1}^{[r+1]} \over t_{i+r}^{[r']}}\right)
\ln\left({t_{i-1}^{[r+1]} \over t_{i-1}^{[r+r'+1]}}\right)
\right. \cr  + & \left. {1\over 2}\ln^2 \left(
{t_{i-1}^{[r+1]}\over t_{i}^{[r+r']}} \right)+  {\rm Li}_2
\left( 1- {t_{i}^{[r]}\over t_{i-1}^{[r+1]}} \right) + {\rm Li}_2
\left( 1- {t_{i-1}^{[r+r'+1]}\over t_{i}^{[r+r']}} \right) - {\rm
Li}_2 \left( 1- {t_i^{[r]}t_{i-1}^{[r+r'+1]}\over
t_{i-1}^{[r+1]}t_{i}^{[r+r']}} \right)+\CO(\epsilon)\right]. }}
The dilogarithm function is defined by ${\rm Li}_2(x) = -\int_0^x \ln
(1-z)dz/z$.

\eqn\expi{\eqalign{I^{4m} &= {1\over a (x_1-x_2)}\sum_{j=1}^2 (-1)^j
\left( -\half \ln^2(-x_j) \right. \cr &   -{\rm Li}_2\left( 1+
{\K_{34}-i\epsilon \over \K_{13}-i\epsilon }x_j \right)
-\eta\left( -x_k, {\K_{34}-i\epsilon \over \K_{13}-i\epsilon }
\right) \ln \left( 1+ {\K_{34}-i\epsilon \over \K_{13}-i\epsilon
}x_j \right) \cr & -{\rm Li}_2\left( 1+ {\K_{24}-i\epsilon \over
\K_{12}-i\epsilon }x_j \right) -\eta\left( -x_k,
{\K_{24}-i\epsilon \over \K_{12}-i\epsilon } \right) \ln \left( 1+
{\K_{24}-i\epsilon \over \K_{12}-i\epsilon }x_j \right) \cr &
\left. +\ln(-x_j)(\ln(\K_{12}-i\epsilon ) + \ln(\K_{13}-i\epsilon
) - \ln(\K_{14}-i\epsilon ) - \ln(\K_{23}-i\epsilon )  ) \right). }
}

Here we have defined $\K_{ml} \equiv -(K_m + K_{m+1} + \ldots +
K_{l-1})^2$.

\eqn\etabranch{\eta(x,y)=2\pi i [\vartheta(-\Im x)\vartheta(-\Im
y)\vartheta(\Im(xy)) -\vartheta(\Im x)\vartheta(\Im
y)\vartheta(-\Im(xy))], }
and $x_1$ and $x_2$ are the roots of a quadratic polynomial:
\eqn\quoio{a x^2+b x+ c + i\epsilon d = a (x-x_1)(x-x_2),}
with
\eqn\defss{\eqalign{ &  a = \K_{24}\K_{34}, \cr & b=
\K_{13}\K_{24}+ \K_{12}\K_{34}-\K_{14}\K_{23}, \cr & c =
\K_{12}\K_{13}, \cr & d = \K_{23}. }}
%

\appendix{B}{Tree Amplitudes with Fermions and Scalars}

Here we summarize some results that are useful for our calculations.\foot{Some of these results have appeared in \refs{\LuoRX,\LuoMY}.}
Take $a=2$ for a scalar and $a=1$ for a fermion.  (Taking $a=0$ reproduces the results for all-gluon amplitudes, but these are not needed in this paper.)

\eqn\lageno{\eqalign{
 &  A( 4^+_{F/S}, 5^+, 6^+,1^-,2^-,3^-_{F/S}) = I_{12|3456}
+I_{1234|56} \cr
& =  { \gb{3|1+2|6}^3\over [6~1][1~2]\braket{3~4}\braket{4~5}
P_{345}^2 \gb{5|6+1|2}} \left( -{ \gb{4|1+2|6}\over \gb{3|1+2|6}}\right)^a
\cr
& + { \gb{1|5+6|4}^3\over [2~3][3~4]\braket{5~6}\braket{6~1} P_{561}^2
\gb{5|6+1|2}} \left({ \gb{1|5+6|3}\over \gb{1|5+6|4}}\right)^a
}}

Similarly we have
\eqn\golfi{\eqalign{
 &  A( 4^-_{F/S}, 5^+, 6^+,1^-,2^-,3^+_{F/S}) = I_{12|3456}+
+I_{1234|56} \cr
& =  { \gb{4|1+2|6}^4\over [6~1][1~2]\braket{3~4}\braket{4~5}
P_{345}^2 \gb{5|6+1|2}\gb{3|1+2|6}} \left({ \gb{3|1+2|6}
\over \gb{4|1+2|6}}\right)^a\cr
&  + { \gb{1|5+6|3}^4\over [2~3][3~4]\braket{5~6}\braket{6~1} P_{561}^2
\gb{5|6+1|2}\gb{1|5+6|4}} \left(-{\gb{1|5+6|4}\over \gb{1|5+6|3}} \right)^a
}}

Notice that for the split $I_{123|456}$ with
a fermionic line between the two vertices, the amplitude is zero
(unlike the internal gluon case).

\eqn\phasco{\eqalign{
&  A(1^+_{F/S},2^-,3^+,4^+, 5^-, 6^-_{F/S})=
I_{12|3456}+I_{612|345}+I_{5612|34} \cr
& =  {  [1~3]^4\braket{5~6}^4 \over [1~2][2~3]\braket{4~5}\braket{5~6}
P_{123}^2 \gb{4|P_{123}|1} \gb{6|P_{123}|3}}
\left( -{\gb{5|P_{123}|3}\over [1~3]\braket{5~6}}\right)^a\cr
&  + { \braket{6~2}^4 [3~4]^4 \over \braket{6~1}\braket{1~2}[3~4][4~5]
\gb{2|P_{612}|5} \gb{6|P_{612}|3} P_{612}^2}
\left(- { \braket{2~1}\over \braket{2~6}}\right)^a \cr
&  + { \gb{2|P_{234}|1}^4 \over \braket{2~3}\braket{3~4}[5~6][6~1]
P_{234}^2 \gb{4|P_{234}|1} \gb{2|P_{234}|5}} \left( -
{\gb{2|P_{234}|6}\over \gb{2|P_{234}|1}} \right)^a
}}

\eqn\onchne{\eqalign{
&  A(1^-_{F/S},2^-,3^+,4^+, 5^-, 6^+_{F/S})= I_{23|4561}+ I_{123|456}
+I_{6123|45} \cr
& =  { [3~4]^4 \braket{5~1}^4 \over [2~3][3~4]
\braket{5~6}\braket{6~1} P_{234}^2 \gb{1| P_{234}|4} \gb{5| P_{234}|2}}
\left( {\braket{5~6}\over \braket{5~1}}\right)^a \cr
&  + { \braket{1~2}^4 [4~6]^4 \over \braket{1~2}\braket{2~3}
[4~5][5~6] \gb{3|P_{456}|6} \gb{1|P_{456}|4} P_{456}^2}
\left( {  \gb{2|P_{456}|4} \over \braket{2~1}[6~4]}\right)^a\cr
&  + { \gb{5| P_{345}|6}^4 \over \braket{3~4}\braket{4~5}
[6~1][1~2] P_{345}^2 \gb{5| P_{345}|2}\gb{3| P_{345}|6}}
\left( -{ \gb{5| P_{345}|1} \over \gb{5| P_{345}|6}}\right)^a
}}

\eqn\neph{\eqalign{
&  A(1^-_{F/S},2^-,3^+,4^-, 5^+, 6^+_{F/S})=
I_{12|3456}+I_{612|345}+I_{5612|34} \cr
& =  { \gb{4|P_{123}|3}^4 \over [1~2][2~3]\braket{4~5}\braket{5~6}
P_{123}^2 \gb{4|P_{123}|1} \gb{6|P_{123}|3}}
\left( { [1~3]\braket{4~6}\over \gb{4|P_{123}|3}}\right)^a\cr
&  + { \braket{1~2}^4 [3~5]^4 \over \braket{6~1} \braket{1~2}[3~4][4~5]
\gb{2|P_{612}|5} \gb{6|P_{612}|3} P_{612}^2}
\left( { \braket{2~6}\over \braket{2~1}}\right)^a \cr
&  + { [5~6]^4 \braket{4~2}^4 \over \braket{2~3}\braket{3~4}[5~6][6~1]
P_{234}^2 \gb{4|P_{234}|1} \gb{2|P_{234}|5}} \left( -
{[5~1]\over [5~6] } \right)^a
}}

\eqn\ysan{\eqalign{
&  A(1^+_{F/S},2^-,3^+,4^-, 5^+, 6^-_{F/S})=
I_{12|3456}+I_{612|345}+I_{5612|34} \cr
& =  {[1~3]^4\braket{4~6}^4 \over [1~2][2~3]\braket{4~5}\braket{5~6}
P_{123}^2 \gb{4|P_{123}|1} \gb{6|P_{123}|3}}
\left(- { \gb{4|P_{123}|3}\over  [1~3]\braket{4~6}}\right)^a\cr
&  + { \braket{6~2}^4 [3~5]^4 \over \braket{6~1} \braket{1~2}[3~4][4~5]
\gb{2|P_{612}|5} \gb{6|P_{612}|3} P_{612}^2}
\left( -{ \braket{2~1}\over \braket{2~6}}\right)^a \cr
&  + { [5~1]^4 \braket{4~2}^4 \over \braket{2~3}\braket{3~4}[5~6][6~1]
P_{234}^2 \gb{4|P_{234}|1} \gb{2|P_{234}|5}} \left(
{[5~6]\over [5~1] } \right)^a
}}

The NMHV tree-level amplitude  with adjacent negative helicities is given by
\eqn\eesixteenone{\eqalign{
 &  A(4_{F/S}^+,5^+,...,n^+,1^-,2^-,3^-_{F/S})\cr
 = &
 -{1 \over{\prod_{i=3}^n \vev{i~i+1}}}
\sum_{j=0}^{n-5}
{ \braket{n-j-1~~n-j} \braket{1| K_2^{[n-j-2]}K_3^{[n-j-3]}|3}^3
\over  t_2^{[n-j-2]} t_3^{[n-j-3]} \gb{n-j|K_2^{[n-j-2]}|2}
\gb{n-j-1|K_2^{[n-j-2]}|2}} \cr & \times
\left( -{\braket{1|  K_2^{[n-j-2]}
K_3^{[n-j-3]}|4} \over \braket{1|  K_2^{[n-j-2]}
K_3^{[n-j-3]}|3} }\right)^a
}}


\listrefs

\end